# Reduced-order particle-in-cell simulations of a high-power magnetically shielded Hall thruster

M. Reza[*1], F. Faraji[*], A. Knoll[*], A. Piragino[†2], T. Andreussi[†2], T. Misuri[‡]

[*] Plasma Propulsion Laboratory, Department of Aeronautics, Imperial College London, London, United Kingdom

[†] Department of Civil and Industrial Engineering, University of Pisa, Pisa, Italy

[‡] SITAEL SpA, Pisa, Italy

**Abstract**: High-power magnetically shielded Hall thrusters have emerged in recent years to meet the needs of the next-generation on-orbit servicing and exploration missions. Even though a few such thrusters are currently undergoing their late-stage development and qualification campaigns, many unanswered questions yet exist concerning the behavior and evolution of the plasma in these large-size thrusters that feature an unconventional magnetic field topology. Noting the complex, multi-dimensional nature of plasma processes in Hall thrusters, high-fidelity particle-in-cell (PIC) simulations are optimal tools to study the intricate plasma behavior. Nonetheless, the significant computational cost of traditional multi-dimensional PIC schemes renders simulating the high-power thrusters without any physics-altering speed-up factors unfeasible. The novel reduced-order "quasi-2D" PIC scheme enables a significant reduction in the computational cost requirement of the PIC simulations. Thus, in this article, we demonstrate the applicability of the reduced-order PIC for a cost-efficient, self-consistent study of the physics in high-power Hall thrusters by performing simulations of a 20 kW-class magnetically shielded Hall thruster along the axial-azimuthal and radial-azimuthal coordinates. The axial-azimuthal quasi-2D simulations are performed for three operating conditions in a rather simplified representation of the thruster's inherently 3D configuration. Nevertheless, we have resolved self-consistently an unprecedented 650 µs of the discharge evolution without any ad-hoc electron mobility model, capturing several breathing cycles and approximating the experimental performance parameters with an accuracy of 70 to 80 % across the operating conditions. The radial-azimuthal simulations, carried out at three cross-sections corresponding to different axial locations within the discharge channel, have casted further light on the evolution of the azimuthal instabilities and the resulting variations in the electrons' cross-field mobility and the plasma-wall interactions. Particularly, we observed the development of a long-wavelength, relatively low-frequency wave mode near the exit plane of the thruster's channel that induces a notable electron transport and a significant ion heating.

**Section 1: Introduction**

Today, innovative near-Earth and interplanetary space missions are being planned with the aim of enabling sustainable and responsible utilization of space, on the one hand, and to expand humanity's capability in exploring space, from the Moon to the Mars and beyond, on the other. High-power magnetically shielded Hall thrusters [1][2][3][4] are currently seen as one of the most promising candidates among the electric propulsion (EP) concepts to realize these novel mission scenarios. Indeed, these thrusters provide both the high-throughput and the long-lifetime capabilities that are essential for the next-generation missions [5]. Additionally, these advanced Hall thrusters feature the same high thrust-to-power ratio and operational versatility characteristics that are commonly associated with the Hall thruster technology.

Nevertheless, the now-about-a-decade-long experience of developing and testing this class of Hall thrusters has casted light, on the one hand, on the enormous cost and time required to mature this new technology to the qualified and flight-ready status. On the other hand, extensive research into the Hall thrusters in the past years have demonstrated the influence of the on-ground vacuum testing facilities on the performance and stability of Hall devices in both magnetically shielded and conventional "unshielded" configurations [6][7][8][9]. In fact, the testing environment is observed to alter the underlying physical processes, such as the electrons' cross-field mobility and the global discharge dynamics [10][11][12], thus, notably affecting the reliability of the characterization/qualification test results with respect to the operation on orbit. Finally, it is becoming globally accepted that the current qualification strategies, centered mainly around single life testing of the Hall propulsion system for a pre-determined operating scenario is not the most reliable approach, as it lacks sufficient statistical data for reliable quantification of the risks and devising proper safety margins from a statistical perspective [13].

The above realizations have once again highlighted the necessity of high-fidelity predictive models that can aid the design, development, and testing of Hall thrusters, thus, mitigating the required cost and time from one side and augmenting the reliability of this technology from the other side, particularly in view of the complex, demanding mission scenarios for which these devices are prime candidates.

---

[1] **Corresponding Author** (m.reza20@imperial.ac.uk)
[2] Former affiliation: SITAEL SpA

In any case, years of effort toward finding reliable and generalizable closure models for the low-computational-cost fluid plasma simulation codes [14]-[19] have not yet been fully successful such that these numerical tools can be used as predictive models of the plasma behavior in Hall thrusters. Furthermore, although traditional fully kinetic PIC codes provide a high level of fidelity and have the capacity to reveal the underlying physical processes and their interactions, their computational cost without any physics-altering speed-up scaling factors are so enormous for real-size high-power Hall thrusters that their application as a design and test-aiding tool is impractical. To provide a numerical sense to this significant computational cost, according to Ref. [20], simulating 150 μs of discharge evolution in a medium-power-class 5 kW Hall thruster in the axial-azimuthal configuration would require more than a year on 1500 CPUs. Noting this unaffordable cost of traditional PIC simulations for modeling advanced Hall thrusters, the study of the physics of operation in these devices has been carried out in recent years using hybrid fluid/PIC simulations [21]-[23] that, although have provided valuable insights into plasma phenomena such as the erosion [21], lack a self-consistent description of the electrons' dynamics and, hence, rely on using ad-hoc electrons' transport models that require tuning with experimental data.

In this regard, the reduced-order PIC scheme can serve as a breakthrough in kinetic modeling of the plasma in advanced Hall thrusters by enabling self-consistent simulations at a fraction of the computational cost compared to the traditional multi-dimensional PIC schemes. This indeed is the main objective of the present work to demonstrate the applicability of the reduced-order PIC as a cost-efficient modeling tool for engineering purposes. The concept of the reduced-order PIC scheme and its implementation for quasi-2D simulations as an electrostatic explicit code named IPPL[1]-Q2D are presented in Refs. [24][25]. The reduced-order PIC is predicated on a decomposition of the simulation domain into the so-called "regions" and a dimensionality-reduction technique that enables resolving the variations in the plasma parameters separately along the involved simulation directions within each region. The IPPL-Q2D code is extensively verified in the axial-azimuthal and radial-azimuthal configurations representative of Hall thruster's geometry [24][25]. We have particularly demonstrated that, in the axial-azimuthal coordinates, the "single-region" quasi-2D simulation, which provides two orders of magnitude reduction in the computational cost, captures an average effect of the azimuthal instabilities on electrons' cross-field transport that results in a reasonably accurate prediction of the plasma dynamics along the thruster's axial direction [26][27][28]. Moreover, in the radial-azimuthal coordinates, we have shown that "multi-region" quasi-2D simulations that offer about an order of magnitude speed-up compared to a full-2D simulation provide high-fidelity predictions of the underlying physical processes and the plasma properties' distribution that compare well with the full-2D results [25]. As such, we used the "multi-region" quasi-2D simulation in a previous work [29] to perform a comprehensive investigation into the effects of the magnetic field gradients, the Secondary Electron Emission and the plasma density on the instabilities and the wave-induced electron transport in a generic radial-azimuthal E × B simulation setup.

Accordingly, we present in this article our results from the single-region axial-azimuthal and the multi-region radial-azimuthal quasi-2D PIC simulations of a development model of a 20 kW-class magnetically shielded Hall thruster, SITAEL's HT20k, which is described in Refs. [12] and [30]. It is worth noting that magnetically shielded Hall thrusters, such as the HT20k, have a complex magnetic field topology with large gradients along the radial direction. This implies that the plasma configuration in these thrusters exhibits a more pronounced 3D nature than the Hall thrusters with conventional magnetic topologies. Accordingly, an axial-azimuthal simulation configuration for the HT20k inherently does not resolve the radial gradients. In any case, by performing the quasi-2D axial-azimuthal simulations using the single-region approximation, we intend to primarily investigate the level of accuracy at which such a self-consistent, low-computational-cost kinetic simulation can predict the experimental performance parameters and the global discharge dynamics. In turn, the radial-azimuthal simulations are carried out to complement the studies along the axial-azimuthal direction by underlining the significance of the radial gradients in the magnetic field on the radial distribution of the plasma properties and the spatiotemporal evolution of the involved plasma processes.

**Section 2: Quasi-2D axial-azimuthal simulations**

The setup of the single-region quasi-2D axial-azimuthal simulations is described in Section 2.1. In Section 2.2, we discuss the simulations' predictions in terms of the global performance parameters and the macroscopic discharge properties as well as the characteristics of the azimuthal waves, the evolution of the plasma species' axial and azimuthal distribution functions, and the overall distribution of the electrons' axial mobility.

---

[1] Imperial Plasma Propulsion Laboratory

**2.1. Simulations' setup and conditions**

The axial-azimuthal simulations are quasi-2D in the configuration space and 3V in the velocity space. The domain of the simulations is a Cartesian $x-z$ plane, with the $x$-coordinate along the axial direction and the $z$-coordinate along the azimuth. The anode is located at the left-hand side of the domain whereas the cathode boundary is at the right-hand end. Along the axial direction, the channel exit plane is located at the middle of the domain, i.e., $x/L_x$ = 1.0. Accordingly, a plume length equal to the channel's axial extent is modelled. In the azimuthal direction, the computation domain is 2-cm-long.

At the beginning of the simulation, the electrons and ions are sampled from a Maxwellian distribution at 10 eV for the electrons and 0.5 eV for the ions. These particles are then injected uniformly throughout the domain at exactly the same locations. In order to maintain the discharge, electrons are sampled from a half-Maxwellian at 5 eV and are injected into the domain at each timestep from the cathode boundary. The number of electrons to inject is determined based on the quasi-neutrality approach at the cathode [20].

As referred to in Section 1, the "single-region" decomposition [24][27] is used to approximate the 2D plasma potential. In IPPL-Q2D, the coupled system of 1D Poisson's equations, obtained by applying the dimensionality-reduction technique, is solved for the 1D potential functions along the simulations' directions using the Reduced-Dimension Poisson Solver (RDPS) [24]. Dirichlet potential boundary conditions are applied at both axial ends of the domain, with the cathode side at 0 V and the anode side at 300 V. In the azimuthal direction, a periodic boundary condition has been applied to either end. The axial and azimuthal potential boundary conditions are applied in the RDPS as described in Ref. [24].

Concerning the boundary conditions for the plasma particles, the azimuthal direction is assumed as periodic. Therefore, the particles leaving the domain at one end are injected back from the other end. Along the axial direction, both ions and electrons reaching either the anode or the cathode side of the domain are removed from the simulation. However, ion recombination on the anode and the consequent injection of the neutrals are considered in the simulations.

The neutrals' dynamics in the axial-azimuthal simulations is resolved in terms of their density and velocity evolution by solving a 1D axial Euler system of fluid equations. The adoption of a fluid description for the neutrals is in line with the current practices being pursued by the community for the axial-azimuthal simulations [20][31]. The details of our neutrals' fluid solver, its benchmark results, and its application in 1D axial and quasi-2D axial-azimuthal simulations of a SPT100-type Hall thruster domain can be found in Ref. [32]. In this reference, the authors demonstrated that, in the absence of the radial dimension and especially the radial expansion of the neutrals in the plume, the use of 1D Euler system of equations for the neutrals can be a reasonable approximation of their kinetic-DSMC treatment.

The flow of the neutrals is assumed to be isothermal at the temperature of 650 K. Thus, the energy equation is not solved, and the Euler system is closed using the equation of state. As the neutrals are treated as a fluid, to account for the neutrals created due to the ion recombination on the anode, the corresponding mass flow rate is calculated and added to the mass flow rate injected from the gas distributor [20]. The effects of the background pressure due to the residual neutrals in the vacuum chamber and the neutrals population in the cathode plume are neglected. Hence, no neutral injection is considered from the cathode boundary.

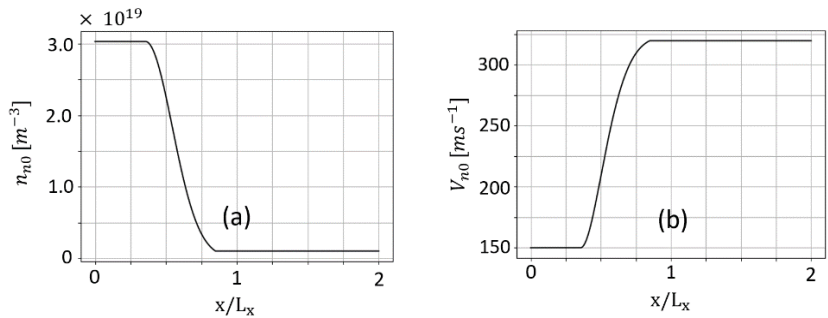

Figure 1: Initial axial profiles of the neutral density (left) and neutral velocity (right) used for the axial-azimuthal HT20k simulations

The initial profiles of the neutral density and axial velocity are shown in Figure 1. It is noted that the specific choice of the initial neutral density and velocity profile mostly affects the initial discharge transient, and simulations with different choices of these initial profiles shall eventually converge to a similar quasi-steady state. Accordingly, the initial profiles of Figure 1 are not expected to notably influence the plasma behavior beyond the initial transient phase of the discharge evolution.

The electron-neutral collisions are resolved in the simulations using the Monte-Carlo Collisions (MCC) scheme [33]. The following collision events are included in the MCC module for the present study: the single ionization, four excitations and the elastic momentum-transfer collision. The cross-sections for these collisions are taken from the Biagi-v7.1 Dataset [34]. The ion-neutral and ion-electron collisions are not considered. Particularly concerning the e-i collisions, prior works, for instance Ref. [20], have found their effect to be minor especially when not accounting for electron-wall interactions.

Table 1 presents a summary of the numerical and physical parameters used for the quasi-2D axial-azimuthal simulations.

| Parameter | Value [unit] |
|---|---|
| **Computational parameters** | |
| Cell size ($\Delta x = \Delta z$) | 20 [$\mu m$] |
| Time step ($ts$) | $2 \times 10^{-12}$ [$s$] |
| Initial number of macroparticles per cells for axial grid ($N_{ppc,\ x}$) | 25 |
| Initial number of macroparticles per cells for the azimuthal grid ($N_{ppc,z}$) | 120 |
| Total simulated time | 650 [$\mu s$] |
| **Physical parameters** | |
| Initial plasma density ($n_{i,0}$) | $1 \times 10^{17}$ [$m^{-3}$] |
| Electron temperature ($T_{e,\ 0}$) (initial load) | 10 [$eV$] |
| Electron temperature ($T_{e,\ c}$) (cathode injection) | 5 [$eV$] |
| Ion injection temperature ($T_{i,0}$) | 0.5 [$eV$] |
| Neutral temperature ($T_n$) | 650 [$K$] |
| Anode voltage ($V_a$) | 300 [$V$] |

Table 1: Summary of the numerical and physical parameters used for the HT20k axial-azimuthal simulations

The axial-azimuthal simulations are performed for three operating conditions of the HT20k where the thruster was operated on xenon as the propellant. An overview of the operating parameters is provided in Table 2. The values of the peak magnetic field intensity for the different operating conditions are normalized with respect to the peak field intensity of Case II, which is denoted as $B_0$.

| Simulation case/ Operating parameter | Discharge voltage [V] | Anode mass flow rate [mg/s] | Magnetic field peak intensity [mT] |
|---|---|---|---|
| I | 300 | 25 | $B_1 = 0.8B_0$ |
| II | 300 | 30 | $B_2 = B_0$ |
| III | 300 | 35 | $B_3 = 1.5B_0$ |

Table 2: HT20k operating conditions used for the quasi-2D axial-azimuthal simulations

The axial profiles of the normalized radial magnetic field intensity for various simulated operating conditions are shown in Figure 2. For the axial-azimuthal simulations, the magnetic field has only a radial component (outward from the simulation plane) and its magnitude along the axial direction varies according to Figure 2.

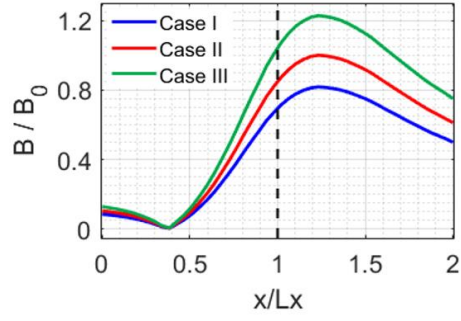

Figure 2: Axial profiles of the normalized radial magnetic field intensity for the three simulated cases

## 2.2. Results and discussion

We start discussing the results from the quasi-2D axial-azimuthal simulations by referring to Table 3 and Figure 3. Table 3 shows the predicted performance parameters for each operating condition whereas, in Figure 3, we have compared the variation vs the anode mass flow rate (AMFR) of the simulated performance and the measured one as reported in Ref. [12]. The data are shown in a non-dimensional format in the figure at the request of SITAEL. In this regard, each predicted/measured performance parameter is normalized with respect to the corresponding measured value of that parameter for Case II with the anode mass flow rate of 30 mg/s.

As mentioned in Section 2, we have not considered the neutrals' background pressure effect in the simulations and, as a result, our setup amounts to a thruster operating in space where the background pressure is indeed negligible. Accordingly, for the three simulated operating conditions, we have provided in Figure 3 the experimental performance at a higher and a lower background pressure level. In this regard, the experimental datapoints illustrated as green dots in Figure 3 correspond to the performance measurements at a reduced background pressure by a factor of 2.

| | Simulation | | |
|---|---|---|---|
| | **Case I** | **Case II** | **Case III** |
| $I_d$ [$A$] | 22.8 | 27.5 | 29.8 |
| $T$ [$mN$] | 350 | 456 | 510 |
| $I_{sp}$ [$s$] | 1906 | 1834 | 1773 |
| $\eta_T$ [%] | 35.8 | 42 | 41.6 |

Table 3: Time-averaged performance parameters from the quasi-2D simulations of the HT20k in the three operating conditions

The predicted performance values shown in Table 3 are averaged over several breathing cycles. In this respect, the instantaneous value of the discharge current ($I_d$) is calculated in our simulations as the difference between the electron and ion flux reaching the anode at each time step. Furthermore, we have estimated the instantaneous values of thrust ($T$) and specific impulse ($I_{sp}$) during the simulation using the following relations:

$$T = \frac{A_{real}}{A_{sim}} \sum_n \mathcal{W}_n M_i v_{ix,n}, \qquad \text{(Eq. 1)}$$

$$I_{sp} = \frac{\sum_n \mathcal{W}_n v_{ix,n}}{g_0 \sum_n \mathcal{W}_n}. \qquad \text{(Eq. 2)}$$

In Eqs. 1 and 2, $A_{real}$ and $A_{sim}$ are the real and simulated thruster's cross-sectional area, respectively, $\mathcal{W}_n$ is the macroparticle weight of each ion crossing the cathode boundary at each timestep, $v_{ix,n}$ is the axial velocity of each exiting ion, $M_i$ is the xenon ion mass, and $g_0$ is the gravitational acceleration at the sea level. The thrust efficiency is calculated using the time-averaged values of thrust and discharge current as

$$\eta_T = \frac{T^2}{2\dot{m}_n I_d V_a}, \qquad \text{(Eq. 3)}$$

in which, $\dot{m}_n$ is the injected xenon mass flow rate from the anode.

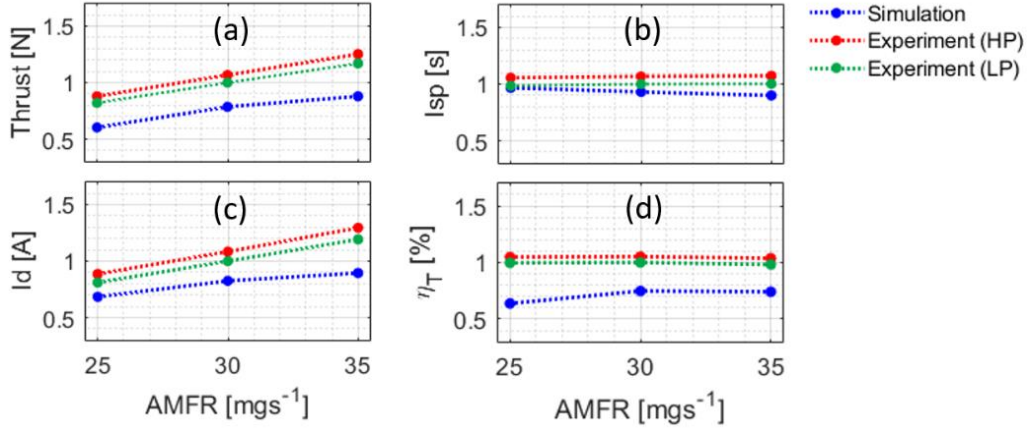

Figure 3: Variation vs the anode mass flow rate of the normalized global performance parameters from the simulations and the experiments [12]; (a) Thrust, (b) Specific impulse, (c), Discharge current, (d) Thrust efficiency. The displayed lower-pressure (LP) data points correspond to the performance measured at a background pressure lowered by a factor of 2 compared to the higher-pressure (HP) points.

From Figure 3, we observe that, as the mass flow rate increases from Case I to Case III, the predicted time-averaged thrust and discharge current show an increasing trend that is consistent with the experimental trend. The thrust efficiency also exhibits an almost similar variation with the mass flow rate to that from the experiments.

However, the specific impulse from the simulations is seen to slightly decrease with increasing mass flow rate whereas it is almost constant in the experiments. The reason for this different trend is that the simulations have predicted a relatively notable population of slow ions in the plume which became more significant at higher mass flow rates. This explanation will be better clarified later when we present the distributions of the macroscopic plasma properties. Nevertheless, the existence of such slow ion population in the axial-azimuthal simulations, which has been most likely absent in the plume of the real thruster, is believed to be due to the fact that the axial-azimuthal simulations do not resolve the confining effect of the radial magnetic field gradient inherent to the magnetic shielding topology that exist in reality. This lack of resolving the radial gradients will hence act as a source of discrepancy between the predicted thrust and Isp from the simulations vs. their measured values. In fact, as it will be confirmed from the radial-azimuthal simulations' results in Section 3.2, the radial gradient of the magnetic field affects the plasma processes regulating the electrons' energy. Hence, in case the radial coordinate had been resolved, an enhancement of the ionization inside the channel could be expected which would have resulted in most of the ions to be created ahead of the region of maximum electric field intensity (the acceleration zone), thus, reducing notably the slow ion population.

Referring to Figure 3, the quasi-2D simulations in all simulated Cases provide self-consistent predictions of the performance that differ from the experimental values at the reduced backpressure by the error ranges of 15-25 % in the discharge current, 20-25 % in the thrust, and 2-10 % in the specific impulse. The error in the thrust efficiency is also between 25-30 % for all Cases compared to the experimental $\eta_T$.

The errors in the performance parameters are deemed to be reasonably acceptable across all simulated Cases, especially considering the simple representation of the actual thruster in the axial-azimuthal simulations and the simplifying assumptions made. However, we point out below, in the order of significance, the main factors that we consider as the root causes of the discrepancies observed between the experimental and simulated performance. It shall be noted that the predicted performance parameters are expected to converge toward the experimental ones when all factors below are addressed together in the simulations.

   a. **Lack of resolving the radial physics**: As it was emphasized in Section 1, the magnetic topology and, thus, the plasma configuration in magnetically shielded Hall thrusters is inherently 3D. Therefore, the radial processes and gradients are expected to play a notable role in determining the state of the plasma and the underlying physical phenomena, hence, affecting the global behavior and performance of the device. The radial-azimuthal HT20k simulations of Section 3 were indeed motivated by the above hypothesis, and the results confirmed the major influence that the radial gradients and interactions have on the plasma processes. Accordingly, we believe that the absence of the radial coordinate in the axial-azimuthal simulations and, hence, not resolving the corresponding effects on the dynamics of the system is the most influential source of deviation between the simulated and the measured performance.

b. **Single-region approximation of the 2D problem**: Although the single-region approximation of an axial-azimuthal simulation domain has been demonstrated to self-consistently capture an average effect of the azimuthal instabilities on the electrons' transport, thus, obviating the need to use any ad-hoc mobility models, it is the simplest approximation that can be established for a 2D problem. This implies that, because of the applied degree of order reduction, there can be an inherent error associated with the predictions of the single-region simulations. Order-reduction errors are typical to any reduced-order model, even full-fluid or hybrid simulations, in which the fluid description of the electrons is a reduced-order model in lieu of the more accurate kinetic description that limits the ability of these simulations to resolve certain processes such as the kinetic phenomena affecting the electrons' behavior and/or VDF. In any case, a single-region reduced-order simulation provides the maximum computational advantage over a traditional 2D PIC simulation. This has been the main enabler of this effort to simulate the extended domain of the HT20k thruster over a significant duration without using any physics-altering speed-up factors, which themselves also introduce an error in the simulation's predictions. Of course, according to our previous studies [26][28], although the single-region axial-azimuthal simulations are reasonably accurate, increasing the approximation order by decomposing the domain into multiple regions, especially along the axial direction, can mitigate further potential errors in the predicted performance associated with the order reduction. In this respect, the convergence of the simulation results with increasing order of approximation to a corresponding full-2D PIC simulation has been verified in benchmark cases [24][25], where the corresponding computational speedup associated with each approximation level are as well provided.

c. **Absence of the background pressure and the additional neutrals' flow rate from the cathode**: The background neutrals in the testing facility and the cathode mass flow fraction, that determines the significance of the neutrals' population in the cathode plume, are well known to affect the performance of Hall thrusters [12][35]. These two factors, which are shown to have a similar effect on the discharge dynamics [35], increase the number density of the neutrals in the near-plume of Hall thrusters. The elevated neutral number density is, in part, ingested into the thruster's discharge channel, artificially increasing the effective thruster's mass flow rate [12]. The background and cathode neutrals also modify the global behavior of the thrusters by affecting the underlying processes such as the electrons' cross-field mobility [7][9][12]. Accordingly, we observe from Figure 3 that all measured performance parameters were lower at a reduced vacuum chamber pressure, and that the difference between the performance at higher and lower backpressures slightly increases with increasing mass flow rate. In this regard, despite some efforts in recent years [12][36][37] aimed at establishing relations to quantify the impact of the background pressure on Hall thrusters' performance and stability so as to ultimately deduce the on-orbit performance metrics based on the on-ground characterization results, a generalizable approach/formulation for this purpose is still not available. Therefore, we chose to exclude the effect of neutrals' backpressure in the axial-azimuthal simulations of this paper, acknowledging that it will inherently introduce an error in the predicted performance.

d. **Neglecting the multiply charged ions**: The simulated cases correspond to the operating conditions at a relatively low discharge voltage of 300 V, which justifies the assumption of neglecting multiply charged ions in the simulations. However, in the real-world setting, a population of multiply charged ions, particularly doubly charged, is typically present and contributes to the discharge current, thrust, and specific impulse. As information regarding the percentage of multiply charged ion populations is not available for the studied operating conditions of the HT20k, we are not able to quantify the exact error associated with neglecting these ion populations. However, it is reasonable to state that the lack of resolving multiply charged ions in the simulations might have led to some discrepancy between the simulated and the measured performance.

e. **Absence of ion-neutral charge-exchange (CEX) collisions:** The CEX collisions are known to be significant in the near-plume of Hall thrusters. Moreover, the CEX and the momentum exchange collisions may have a dampening effect on the ion acoustic instability and may also affect the ions' velocity distribution function. However, in Ref. [32], we have shown that, in the axial and axial-azimuthal simulations which do not capture the plume expansion, the inclusion of the CEX can result in underpredicted ions' velocity. This in turn will result in unrealistically high neutrals' temperature and velocity in the plume if the effect of the ion-neutral interactions is included in the neutrals' model. These consequences are indeed inherent to the axial and axial-azimuthal setups which overestimate the CEX collision frequency due to an increased density of the neutrals interacting with the ion beam. As a result, the CEX collisions are not accounted for in the simulations here since an accurate capturing of the CEX collisions and their effects in the absence of resolving the radial coordinate is not feasible.

Following the above remarks, we proceed with the presentation of the axial-azimuthal simulations' results by referring to Figure 4, which shows the time evolution of the discharge current signal for the three simulation cases. In addition, we have plotted in Figure 5 the averaged 1D Fast Fourier Transform (FFT) of the ion number density signals for Cases I to III to assess the frequency content of the axial discharge oscillations.

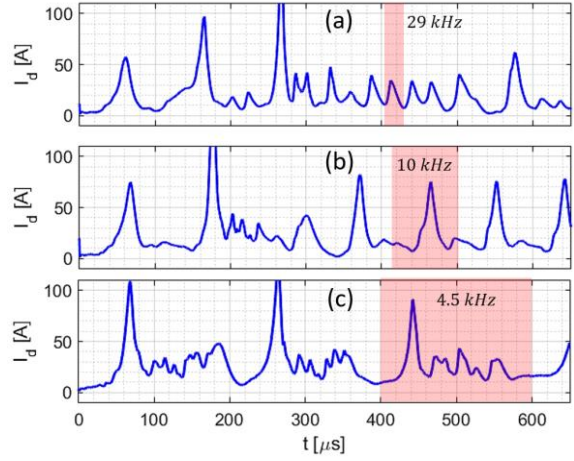

Figure 4: Time evolution of the discharge current for (a) Case I, (b) Case II, and (c) Case III. The highlighted areas isolate one cycle of the periodic behavior in the discharge after reaching the quasi-steady state.

From Figure 4, we notice that all simulations have reached a quasi-steady state and exhibit periodic oscillations reminiscent of the breathing mode. This observation underlines that the average effect of the azimuthal instabilities on the electrons' axial mobility resolved by the single-region quasi-2D simulation has been sufficient to maintain the discharge and to allow the development of the breathing mode oscillations.

Across various operating conditions, however, it is seen in Figure 4 that the amplitude and frequency of the discharge oscillations are different. In this regard, Cases I and III particularly exhibit both "low" and "high"-frequency global discharge oscillations, which are also distinctly evident from the corresponding FFT plots in Figure 5(a) and (c). The peak frequencies predicted by the simulations are consistent with those observed in the experimental discharge current of the HT20k thruster for the three simulated Cases. The observed experimental peak frequencies of the global discharge oscillations have been: Case I, around 9 kHz and 25 kHz, Case II and III: less than 5 kHz and 20 kHz. Even though we cannot eliminate the possibility that the high frequency peaks in the experiments are associated with the cathode operation, they are most likely related to the thruster itself since they were not visible at higher magnetic field intensities.

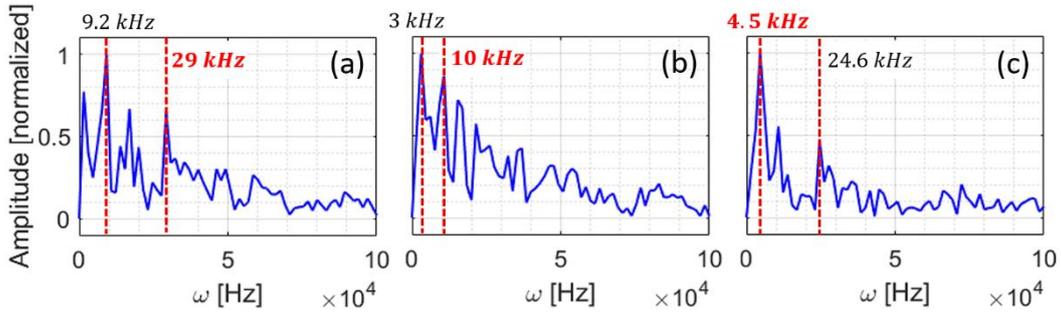

Figure 5: 1D FFT plots of the ion number density signal, averaged over all axial positions, from the quasi-2D simulations; (a) Case I, (b) Case II, and (c) Case III. The two most dominant frequencies in the FFT plots are specified. In each plot, the frequency highlighted in red corresponds to the frequency of the periodic pattern observable in the respective discharge current evolution plot of each Case in Figure 4.

It is observed from Figure 4 that, in Case I, the discharge features a complex evolution with the frequency of the oscillations seen to first increase from about 300 $\mu s$, where the simulation has supposedly reached a quasi-steady state, and then decrease from about 550 $\mu s$. This behavior makes identifying a single breathing frequency rather challenging for Case I.

In Cases II and III, which have increasingly larger values of magnetic field peak intensity and mass flow rates compared to Case I, we see from the plots (b) and (c) in Figure 4 and Figure 5 that the frequency of the dominant repeating cycle in the discharge after reaching the quasi-steady state is about 10 kHz for Case II and 4.5 kHz for Case III. In this regard, the experimental results of Ref. [38] had also interestingly showed a decrease in the main frequency of the global discharge oscillations with increasing magnetic field intensity at discharge voltages around 300 V.

Before proceeding with the discussion of the remainder of the results, we emphasize that the predictions of the axial-azimuthal simulations are not compared against direct intensive measurements of the plasma properties, which are not available for the HT20k. As such, the absence of a detailed experimental validation shall be kept in mind with regard to the results that follow.

Figure 6 to Figure 8 show the spatiotemporal evolution of several macroscopic discharge properties from the quasi-2D simulations of HT20k in the three operating conditions. Figure 6 presents the evolution of the ion number density and the ion axial velocity. Superimposed on each evolution plot of the ion number density in Figure 6(a) is the discharge current signal of that simulated case, which is normalized with respect to the corresponding maximum value of the discharge current. Moreover, the white and cyan curves in Figure 6(b) show, respectively, the variations in the axial position of the points of ions' zero velocity and sonic speed throughout the discharge evolution. It is noted that the point of ions' zero velocity indicates the axial location from where the ions backflow toward the anode.

It is seen in Figure 6(a) that the evolution of the ion number density ($n_i$) is expectedly correlated with the discharge current. In fact, the peaks of the discharge current correspond to the maximum ion (plasma) number density near the anode. In addition, the acceleration and downstream propagation of the ions are observed from the ion axial velocity plots in Figure 6(b). In this regard, the ions created upstream the channel exit are accelerated to velocities of around 20,000 m/s and propagate toward the plume. Accordingly, the location of the points of ions' zero velocity and sonic speed oscillates in time. The axial distance between these two characteristic ion velocity points is also noticed to vary in time such that it becomes maximum when the ions propagate toward the plume and is minimized when the ion front is at the upstream.

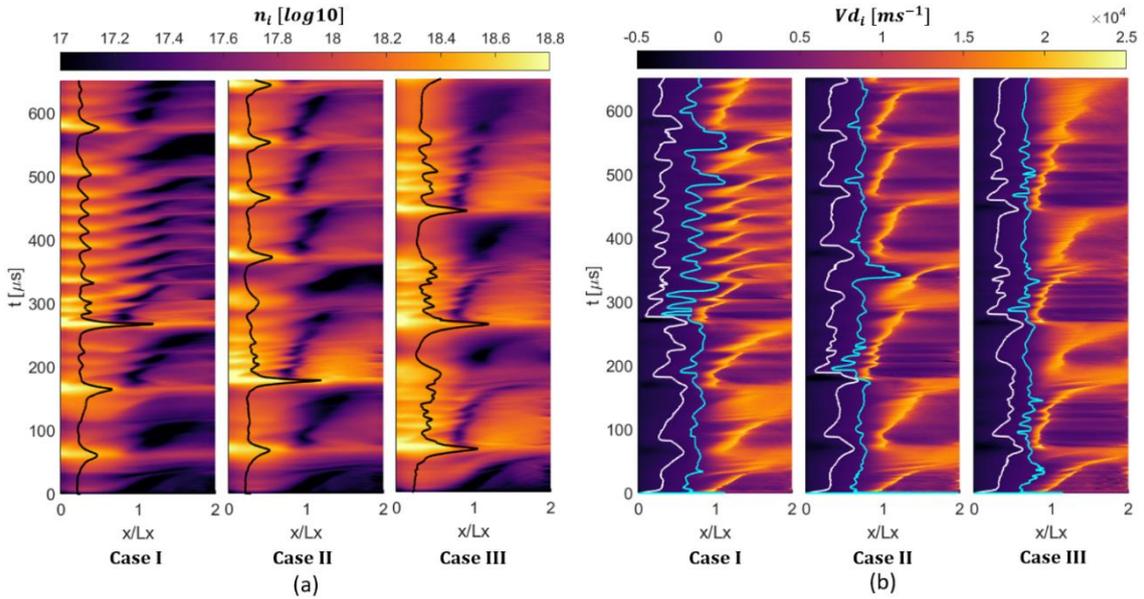

Figure 6: Spatiotemporal maps of (a) ion number density and (b) ion axial drift velocity for the three simulated cases. Superimposed on the ion density plots are the normalized discharge current signals. The white and cyan curves in the ion velocity plots trace the axial position of the points of ions' zero velocity and sonic speed, respectively.

Looking at the evolution plots of $n_i$ and $v_{di}$ for each simulated case, the presence of a low-velocity ion population in the plume can be observed. It is evident that the significance of this slow ion population increases with the increase in anode mass flow rate from Case I to Case III. As we pointed out before when discussing the performance parameters, the drop in the time-averaged ions' exhaust velocity or, equivalently, the $I_{sp}$ from Case

I to Case III is due to the presence of this slow ion population which exists because the ionization within the channel becomes less effective at higher mass flow rates since the important effect of the radial gradients on the axial plasma processes such as the ionization is not resolved in the axial-azimuthal simulations. Hence, with increasing mass flow rate, the ionization occurs over a more extended length scale, leading to the creation of ions that do not "feel" the entire plasma potential drop within the acceleration zone and, therefore, reside in the plume for longer times before exiting the domain.

The spatiotemporal maps of the neutral density ($n_n$) and axial velocity ($V_{dn}$) are presented in Figure 7, which illustrate the resolved dynamics of the neutrals. Referring to Figure 7(a), the neutral density at the anode location ($x = 0$) can be seen to peak for all cases almost immediately following the end of the discharge current drop. This observation can be due to an increase in the ion recombination on the anode as the discharge moves toward the anode when the discharge current decreases. Another noteworthy point concerning the neutral density plots is related to the downstream propagation of the neutrals' front, which can be noticed from the oblique density contours extending from near the anode toward the plume [39].

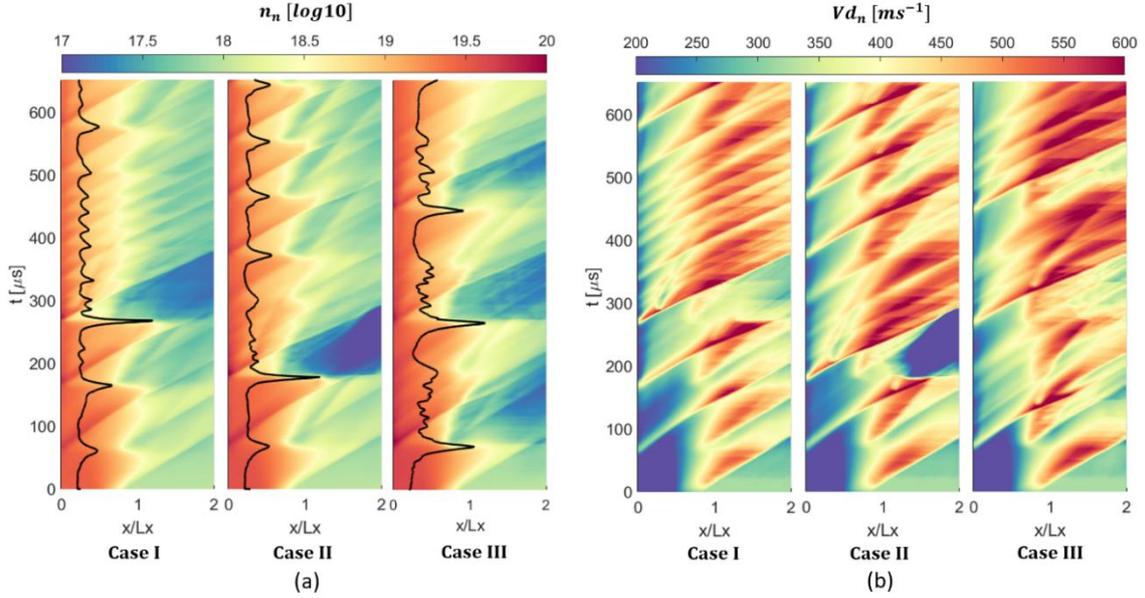

Figure 7: Spatiotemporal maps of (a) neutral number density and (b) neutral axial drift velocity for the three simulated cases. Superimposed on the neutral density plots are the normalized discharge current signals.

Looking at the neutral axial velocity plots in Figure 7(b), we observe that, as the evolution of the neutrals' momentum is resolved in our simulations, the axial velocity of the neutral flow varies from about 200 m/s near the anode to about 600 m/s in the plume. This observed variation in the axial velocity is in line with the experimental observations reported in the literature that suggest a factor of 2 to 3 increase in the axial velocity of the neutrals in Hall thrusters as they propagate downstream [40].

We have shown in the plots (a) and (b) of Figure 8 the spatiotemporal evolutions of the axial electric field ($E_x$) and the electron temperature ($T_e$), respectively. On the electric field plots of each Case, the corresponding normalized discharge current as well as the trace of the axial location of the points of ions' zero velocity and sonic speed are superimposed.

Concerning the electric field evolution (Figure 8(a)), we see that the typical back-and-forth movement of the acceleration zone, represented by the region of maximum electric field intensity, is captured in all cases. In this respect, as reported in the literature [20], the axial electric field peak shifts toward downstream the exit plane ($x/L_x = 1$) during the current increase and retreats inside the channel with the current decrease. It is interesting to note that the extent of the acceleration zone is seen to change in time as well, which is reflected in the electric field plots as a broadening and narrowing of the region of maximum electric field with an associated variation in the peak field's intensity. In line with the movement of the acceleration zone, the axial distance between the points of ions' zero velocity and sonic speed shrinks when the acceleration zone moves upstream and expands when it shifts toward the plume.

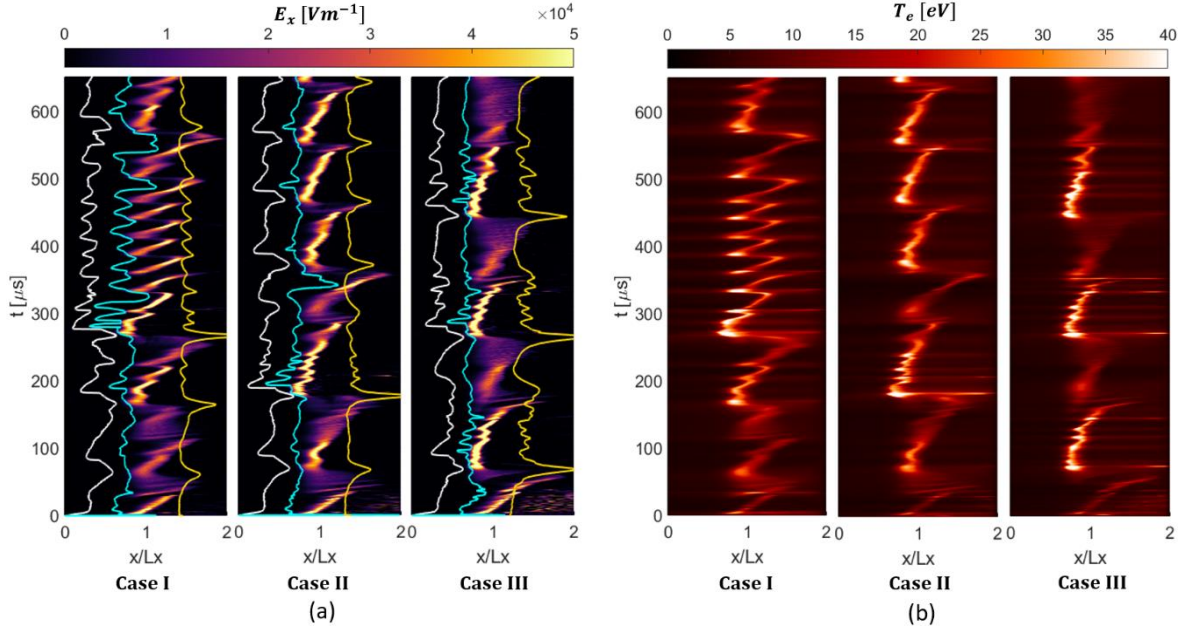

Figure 8: Spatiotemporal maps of (a) axial electric field and (b) electron temperature for the three simulated cases. The yellow curves superimposed on the electric field plots are the normalized discharge current signals. The white and cyan curves trace the axial position of the points of ions' zero velocity and sonic speed, respectively.

Regarding the plots in Figure 8(b), the evolution of the electron temperature is noticed to be completely correlated with that of the electric field. This implies that, as could have been expected, the region of maximum electron temperature corresponds to the region of maximum electric field, i.e., the acceleration zone. Moreover, the electron temperature is seen to reach peak values of over 40 eV in all Cases when the region of maximum electric field intensity is at its narrowest width.

The axial distributions of the discharge properties from the quasi-2D simulations, averaged over 60 to 650 $\mu s$, are plotted in Figure 9 for the three simulated cases. Concerning these time-averaged distributions, it is first observed that, as the magnetic field peak intensity increases from Case I to Case III, so does the gradient of the plasma potential (Figure 9(d)) around the exit plane, illustrated as a dashed back line, resulting in a higher axial electric field peak (Figure 9(b)). In any case, the difference in the peak values of the electric field between Cases II and III is quite minimal. Second, it is noteworthy from the profiles in plots (b) and (h) that, in all Cases, the ions are seen to reach the sound speed, i.e., the Mach number of 1, upstream of the electric field peak. It is noted that the definition of the ions' temperature used for the calculation of the sound speed can be found in Ref. [24].

Third, the increase in the anode mass flow rate from Case I to Case III yields an increase in the ion number density (Figure 9(a)), resulting in a higher neutral density near the anode (Figure 9(e)) due to the combined effects of more significant ion recombination, on the one hand, and an actually higher rate of neutral injection from the gas distributor on the other. Consistent with the rise in the neutral number density near the anode from Case I to III, the ionization rate (Figure 9(c)) also increases. Nevertheless, the difference between the profiles of Cases II and III is again interestingly quite small. No rigorous explanation has been yet found for this observation and it requires further investigation as part of the future work. However, it is emphasized that the small difference between the results of Cases II and III are only in terms of the time-averaged plasma profiles. Indeed, the time evolution of the discharge is distinctly different across all simulated Cases.

In addition, we can observe in Figure 9(c) that the ionization rate profile is rather extended. Because of this extended ionization zone, an ion population is created beyond the peak of the electric field, whose lower axial velocity translates into longer transit times. As a result, the time-averaged profiles in plot (a) of Figure 9 show an increase in the $n_i$ in the plume, which is more pronounced for Cases II and III than for Case I.

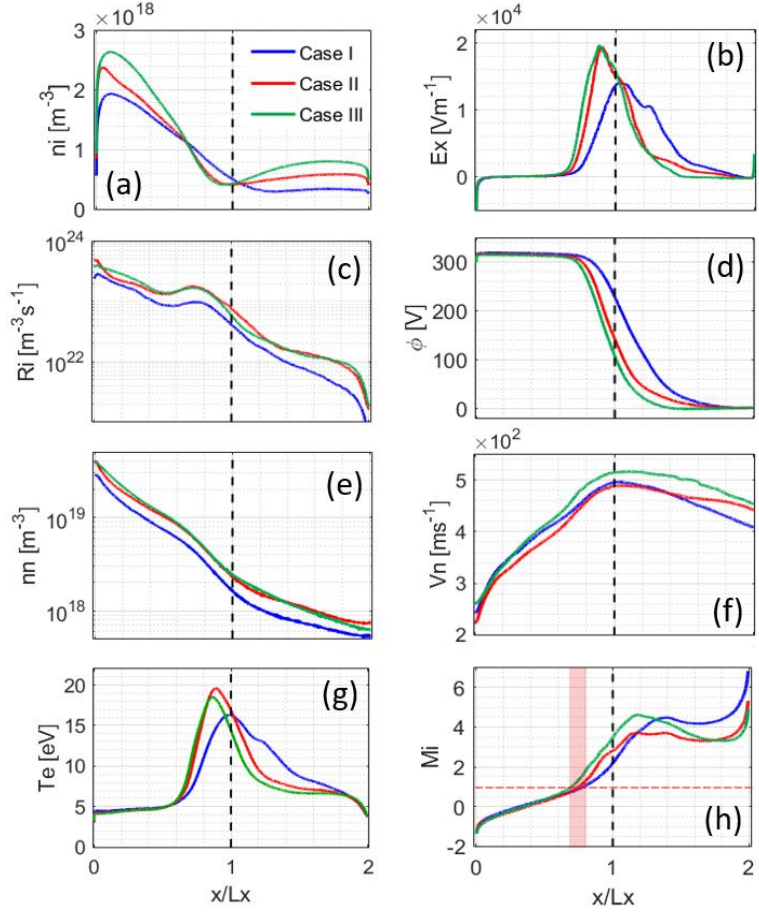

Figure 9: Axial distributions of the time-averaged discharge properties; (a) ion number density, (b) axial electric field, (c) ionization rate, (d) plasma potential, (e) neutral number density, (f) neutral axial velocity, (g) electron temperature, and (h) ion Mach number. In the ion Mach number plot, the shaded area indicates the positions of the ion sonic point.

Fourth, as it was pointed out from the spatiotemporal map of the neutral axial velocity in Figure 7(b), the time-averaged axial profile of the neutral velocity in Figure 9(f) also shows the increase in the axial velocity of the neutral flow from the near-anode zone to the plume.

Finally, concerning the axial profiles of the electron temperature (Figure 9(g)), the time-averaged peaks are consistent with the results reported in Refs. [41][42]. In this regard, it is important to highlight that the simulated cases correspond to relatively low power levels, around 8-12 kW. Since the HT20k is designed to operate at higher power levels (with a nominal condition of 20 kW and 450 V), a lower temperature could be the result of ineffective operation. This is applicable, in particular, to Case I, which shows a lower temperature also in the simulation.

The existence of a slow ion population in the plume is also manifested in the ions' 1D1V distribution function presented in Figure 10 at four different time instances of the discharge evolution for each simulated Case. It is noticed that, at the time of "current low" corresponding to minimum discharge current, a population of ions with very low axial velocities is present beyond the channel's exit plane at $x/L_x = 1$. As the discharge current rises, which is associated with a downstream shift in the region of maximum electric field intensity as observed in Figure 8, the axial velocity of the slow ion population partially increases. Consequently, at the time of "current peak", when the acceleration zone is at its most outward position, a notable dispersion in the ion axial velocity is noticed, which is more pronounced for simulation Cases II and III. During the "current drop", the high velocity ions leave the domain, and the acceleration zone retreats toward upstream. At this time, particularly for Cases II and III, an apparent swirl in the axial velocity of the slow ion population is observed, which is the strongest for Case III. This observation is hypothesized to be due to the trapping of the slow ions by an axially propagating instability wave. Our hypothesis is supported by the fact that prior research in the axial-azimuthal Hall thruster configuration [43] has shown that axial instabilities, such as the Ion Transit Time (ITT), can be excited during the current-decrease phase of the discharge.

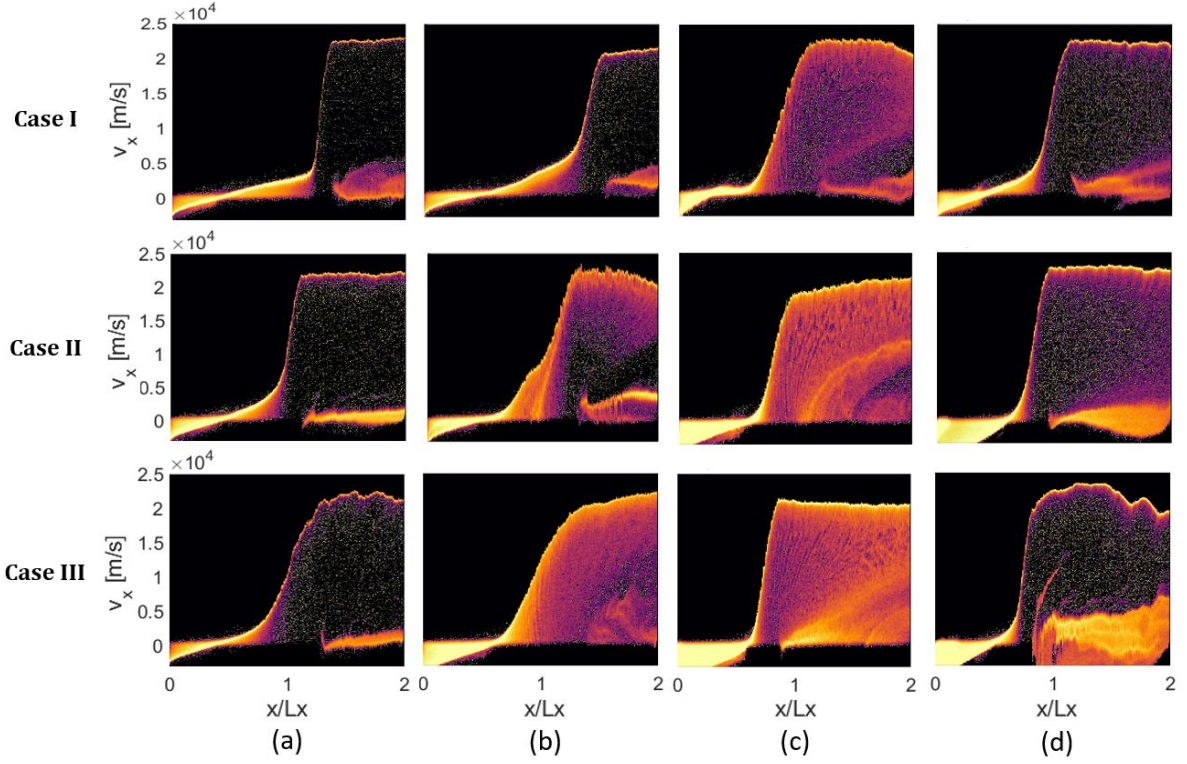

Figure 10: 1D1V velocity distribution functions of the ions along the axial direction and axial velocity component from the quasi-2D simulations of the three operating conditions during different moments of the discharge evolution cycle; (a) current low, (b) current rise, (c) current peak, and (d) current drop.

To investigate the characteristics of the resolved instabilities along the azimuthal direction, we first look at the plots in Figure 11 in which we have shown the 1D FFTs of the azimuthal electric field signal at the mid azimuthal location from the quasi-2D simulations after reaching the quasi-steady state. Moreover, we have highlighted in the FFT plot of each simulated Case the frequency range over which the amplitude of the FFT peaks. It is evident that, across all three operating conditions, the wave modes with frequencies falling in the range of 3-8 MHz are dominant. Accordingly, from the existing literature, we suspect that the frequency peaks are associated with the modified Ion Acoustic Instability which corresponds to the pseudo-saturated state of the Electron Drift Instability (EDI) following its nonlinear growth [44].

To confirm that the dominant wave modes correspond to the Ion Acoustic Instability, we refer to the theoretical dispersion relation of the ion acoustic waves in the laboratory reference frame (Eqs. 4 and 5) [26]. In these equations, $\omega_R$ is the real frequency component of the waves, $\gamma$ is the imaginary frequency component or the growth rate, $k_x$ and $k_z$ are, respectively, the axial and azimuthal wavenumber, $C_s$ is the ion sound speed, $\lambda_D$ is the Debye length, $m_e$ is the electron mass, and $m_i$ is the ion mass.

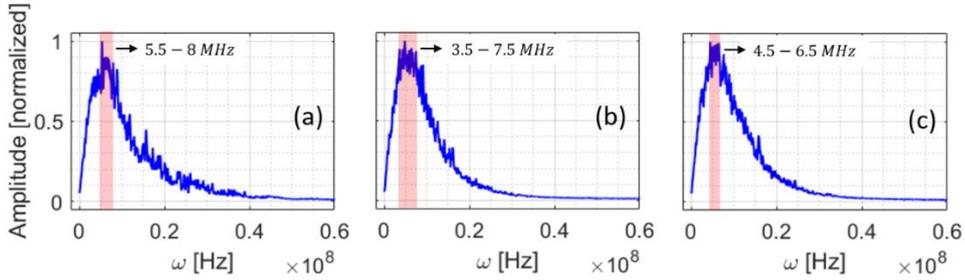

Figure 11: 1D FFT plots of the azimuthal electric field signal at the mid azimuthal location from the quasi-2D simulations of (a) Case I, (b) Case II, and (c) Case III. The shaded areas indicate the frequency contents' peak.

$$\omega_R \approx k_x V_{di,x} \pm \frac{k_y C_s}{\sqrt{1 + k_z^2 \lambda_D^2}}, \qquad \text{(Eq. 4)}$$

$$\gamma \approx \pm \sqrt{\pi \frac{m_e}{m_i} \frac{k_x V_{di,x}}{(1 + k_z^2 \lambda_D^2)^{\frac{3}{2}}}}. \tag{Eq. 5}$$

Following the same analysis performed in Ref. [44], the real frequency and phase velocity ($v_{ph} = \omega/k$) of the mode with the largest growth rate is given by the following relations

$$\omega_{R,max} \approx \frac{C_s}{\lambda_D \sqrt{3}}, \tag{Eq. 6}$$

$$v_{ph,max} \approx C_s \sqrt{\frac{2}{3}}. \tag{Eq. 7}$$

Now, from the time-averaged axial profiles of the discharge properties presented in Figure 9, and recalling the relations for the ion sound speed and the Debye length, namely, $C_s = \sqrt{\frac{eT_e}{M_i}}$ and $\lambda_D = \sqrt{\frac{\epsilon_0 T_e}{en_e}}$, the real frequency and phase velocity of the fastest-growing ion acoustic mode are, respectively, $\omega_{R,max} \approx 10$ MHz and $v_{ph,max} \approx 2$ km/s, assuming the axially averaged values of plasma density ($n_e$) and electron temperature ($T_e$) to be about $1 \times 10^{18}$ $m^{-3}$ and 10 eV across the three simulated cases. The estimated frequency value of 10 MHz from the theory is quite consistent with the frequency range of the dominant modes from the simulations in Figure 11.

Moreover, referring to the time-averaged azimuthal velocity distribution function (VDF) of the ions for the three simulated cases shown in Figure 12, we observe a secondary peak in the VDF around the velocities of 2.5 to 4.5 km/s, which is similar to the theoretical value of the phase velocity obtained above for the fastest-growing ion acoustic wave mode. The appearance of this secondary peak is due to the interactions with the azimuthal waves of the ions whose velocities are near the phase velocity of the wave.

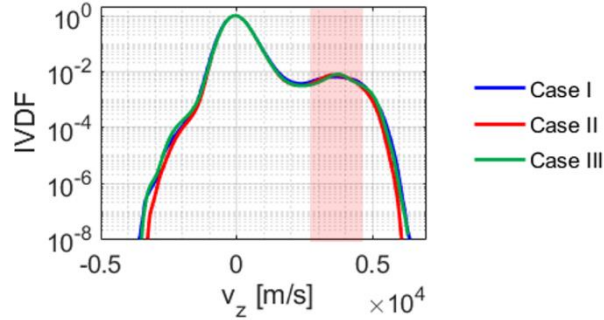

Figure 12: Azimuthal VDF of the ions residing over the entire axial extent of the domain for the three simulated cases. The VDFs are averaged over about 200 $\mu$s of the simulations' time after reaching the quasi-steady state. The shaded area indicates the secondary peak in the VDF associated with the ions interacting with the azimuthal waves.

The time evolution of the characteristics of the azimuthal waves and the numerical dispersion features of the azimuthal electric field fluctuations from the simulations after reaching the quasi-steady state can be discerned by looking at the plots in Figure 13. The plots in the left and middle column of this figure represent, respectively, the spectrogram (or $\omega - t$ diagram) and the time evolution of the 1D spatial FFT of the azimuthal electric field signal for each simulated Case. The normalized discharge current signal is superimposed on the plots of each Case. The plots in the right column correspond to the 2D FFTs of the azimuthal electric field signal from the quasi-2D simulations over the time window of 400 to 500 $\mu$s.

Focusing first on the spectrogram and the 1D spatial FFT plots, it is evident that, in all Cases, the time evolution of the frequency and wavenumber characteristics of the azimuthal waves is perfectly synchronous with the discharge current evolution. In this regard, as the discharge current peaks, the frequency and wavenumber spectra of the waves expand toward higher-frequency, shorter-wavelength modes. In turn, when the discharge current reaches a minimum, the dominant modes feature lower frequencies and longer wavelengths.

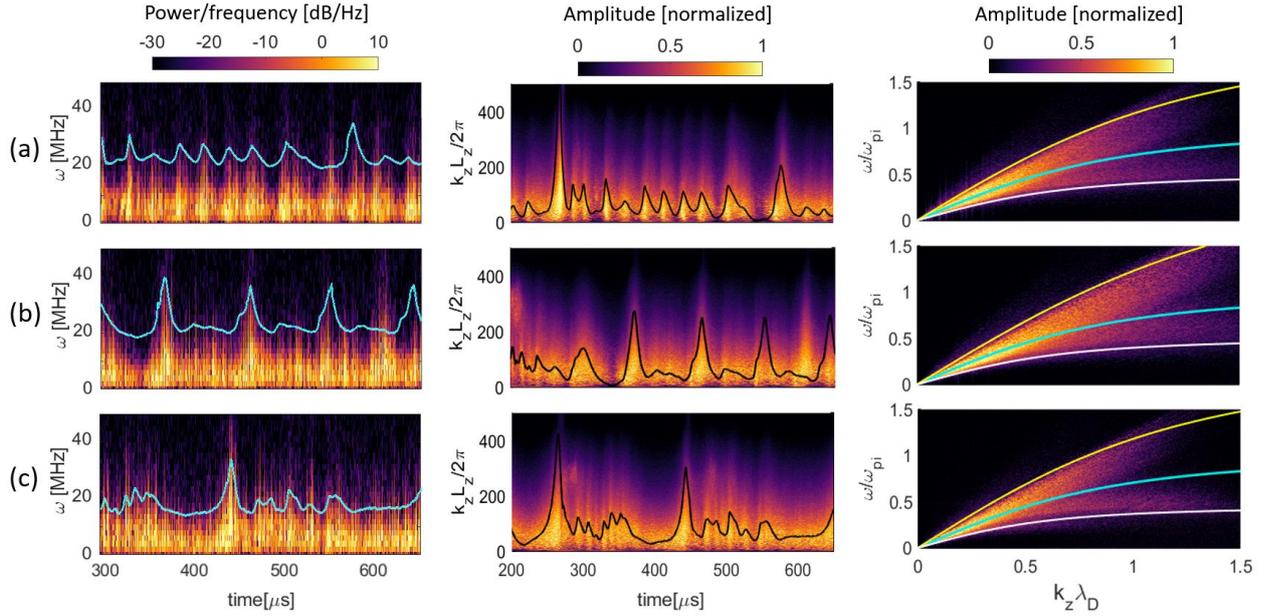

Figure 13: Characteristics of the azimuthal fluctuations for (a) Case I, (b) Case II, and (c) Case III; (left column) spectrogram of the azimuthal electric field signal at the mid azimuthal location over the time window of 300-650 $\mu$s, (middle column) the time evolution within the time interval of 200-650 $\mu$s of the 1D spatial FFT of the azimuthal electric field with the azimuthal wavenumber normalized with respect to the azimuthal domain length, (right column) 2D spatiotemporal FFTs of the azimuthal electric field. The x and y axes in the 2D FFT plots are normalized, respectively, with respect to the ion plasma frequency ($\omega_{pi}$) and the Debye length.

The 2D FFTs of the azimuthal electric field in the right column of Figure 13 also reveal interesting information concerning the characteristics of the azimuthal waves. First, the numerical dispersion of the waves for all three Cases features a rather continuous spectrum over a range of azimuthal wavenumbers ($k_z$). Additionally, for each $k_z$, the modes are seen to span over a certain frequency bandwidth.

Second, the theoretical ion acoustic dispersion relation in the ions' reference frame, i.e., $\omega_R \approx \frac{k_y C_s}{\sqrt{1+k_y^2 \lambda_D^2}}$, which is plotted as a cyan line and using the time-averaged values of the plasma density and the electron temperature over the time duration of 400-500 $\mu$s for each simulated Case, namely, $n_e \sim 1 \times 10^{18}\ m^{-3}$ and $T_e \sim 7$ eV, are very consistent in all simulated Cases with the numerical dispersion of the waves, particularly, for $k_z \lambda_D$ values below about 0.5. In addition, the cyan lines delineate a bifurcation in the numerical dispersion plots, especially for $k_z \lambda_D$ values larger than 1.

Third, the entire 2D FFT spectrum for all three Cases is bounded by the yellow and white curves in Figure 13, which correspond to the theoretical ion acoustic dispersion relation calculated, respectively, using the highest and lowest values of $n_e$ and $T_e$ encountered in each simulation case during the time interval of 400-500 $\mu$s. During this time interval, the $n_e$ and $T_e$ vary significantly by several factors to an order of magnitude depending on each simulated Case. Nonetheless, for all Cases, it is observed that, as the discharge properties evolve in time, the numerical dispersion characteristics of the ion acoustic waves change between two rather distinct states as evident from the bifurcated spectrum of the wave modes in the 2D FFT plots.

To conclude the discussion of the quasi-2D axial-azimuthal simulations' results, we focus on the velocity distribution function and the axial mobility of the electrons. In this regard, Figure 14 shows the electrons' VDF along the axial and the azimuthal velocity components over three axial extents of $0 < x \leq 0.6L_x$, $0.6L_x < x \leq 1.4L_x$, and $1.4L_x < x \leq 2L_x$. Concerning the axial EVDFs in Figure 14(a), we notice that, across all Cases, the VDF deviates most from the Maxwellian within the axial extent of $0.6L_x < x \leq 1.4L_x$, which mostly comprises the acceleration zone. The deviation from the Maxwellian is manifested in terms of the broadening of the distribution function toward the higher velocities.

Regarding the azimuthal VDF (Figure 14(b)), the distribution functions in the three Cases and within all three axial locations feature a side lobe toward positive azimuthal velocities, which is associated with the E × B drift motion of the electrons, whose velocity ($E_x/B_r$) is between 700 to 1000 km/s. This side lobe is also most pronounced within the region of $0.6L_x < x \leq 1.4L_x$ where the electric field intensity is at its peak values.

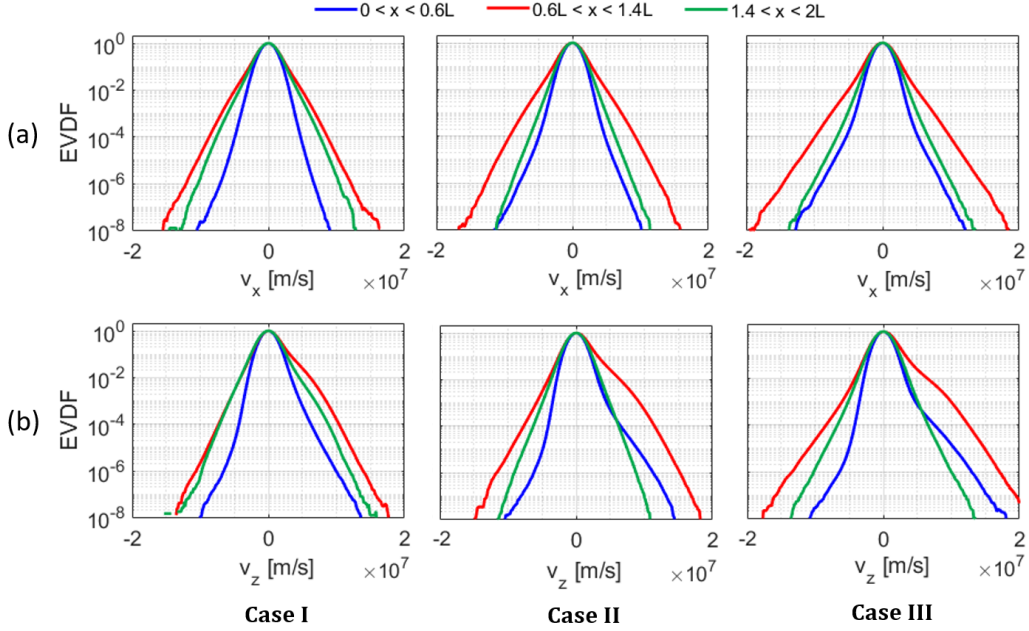

Figure 14: The VDFs of the electrons residing within different axial extents of the domain for the three simulated Cases along (a) axial and (b) azimuthal velocity component. The VDFs are averaged over about 200 $\mu s$ of the simulations' time after reaching the quasi-steady state.

In Figure 15(a), we have shown the time-averaged axial profile of the total electron transport in terms of the magnetic force term ($F_B$), which correspond to the left-hand side of the electrons' azimuthal momentum equation [26] and is defined as

$$F_B = -e \int_{-\infty}^{\infty} v_x B_r f_e(\boldsymbol{v}) d^3v = -e n_e V_{de,x} B_r, \qquad \text{(Eq. 8)}$$

where, $e$ is the elementary charge, $v_x$ is the axial velocity component, $B_r$ is the radial magnetic field intensity, $f_e(\boldsymbol{v})$ is electrons' velocity distribution function, and $V_{de,x}$ is the electron axial drift velocity.

Moreover, Figure 15(b) presents the axial distribution of the equivalent plasma resistivity ($R$) using the relation

$$R = \frac{\bar{E}_x}{\bar{J}_{ex}} = \frac{\bar{E}_x}{-e\bar{n}_e \bar{V}_{de,x}}, \qquad \text{(Eq. 9)}$$

in which, $J_{ex}$ denotes the axial electron current, and the quantities with a bar are time-averaged over 60 to 650 $\mu s$.

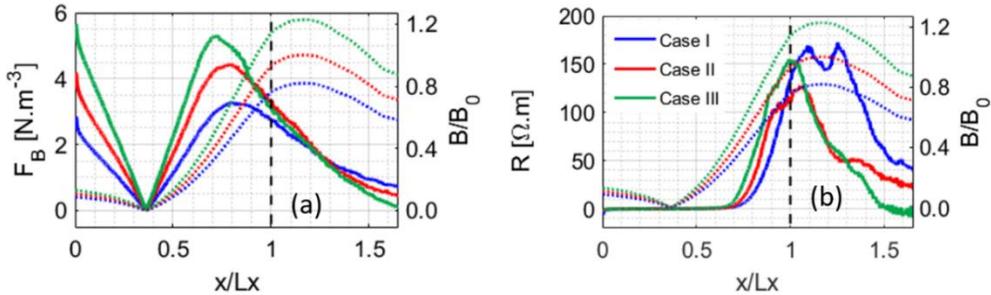

Figure 15: Axial profiles of (a) total electron transport, and (b) equivalent plasma resistivity, averaged over 60 to 650 $\mu s$. The dotted curves are the axial profiles of magnetic field. The dashed black lines show the location of the channel exit.

From Figure 15(a), we observe that the peak of the electron transport in all Cases occurs inside the channel and upstream the peak radial magnetic field intensity. This is consistent with the prior research into the distribution of electrons' axial mobility in Hall thrusters using axial-azimuthal full-2D simulations [20]. Additionally, with the increase in the mass flow rate from Case I to Case III and, consequently, the rise in the plasma number density, the associated stronger wave-particle interactions result in the peak of the electron transport to become larger from Case I to III. Furthermore, corresponding to the location where the magnetic field intensity is roughly zero, i.e., before $x/L_x = 0.5$, the magnetic force term becomes almost zero and it then rises toward the anode side of the domain.

Referring to Figure 15(b), we notice that the electrons expectedly feel the highest resistivity against their axial motion around the location of the peak magnetic field intensity. Nevertheless, for Cases II and III, the maximum resistivity is occurring slightly upstream of the location of peak $B_r$, which is consistent with the time-averaged distribution of the axial electric field for these two Cases shown in Figure 9(b). Moreover, the resistivity is almost zero from $x/L_x$ of about 0.7 toward the anode since the electric field is very small in this region.

### Section 3: Quasi-2D Radial-Azimuthal simulations

The purpose of performing radial-azimuthal simulations of the HT20k in this work is two-fold: (1) to verify the impact that the radial magnetic field gradient in a magnetically shielded Hall thruster has on the radial distribution of the plasma properties and the axial processes such as the electron transport, (2) to cast light on the characteristics and evolution of the azimuthal instabilities in various radial-azimuthal cross-sections of a real-size high-power shielded Hall thruster.

#### 3.1. Simulations' setup and conditions

The quasi-2D radial-azimuthal simulations are carried out using a multi-region decomposition of the simulation domain, which is represented by a Cartesian $y - z$ plane. The $y$-axis is along the radial coordinate and the $z$-axis is along the azimuth. The length of the domain along the azimuthal coordinate is half of that along the radius, whereas the size of the computation cells along the radial and azimuthal directions is selected to be the same and equal to 20 $\mu$m. As a result, for the multi-region decomposition associated with the reduced-order quasi-2D simulations, we have used 50 vertical regions along the radial direction and 25 horizontal regions along the azimuthal direction to keep the ratio of the number of cells per region constant between the two coordinates.

We have simulated three radial-azimuthal cross-sections of the HT20k along the axial ($x$) direction. These three sections and the distribution of the magnetic field properties along the radius in each section are illustrated in Figure 16. From the field topology plot shown in Figure 16(e), it is evident that Section 1 corresponds to the beginning of the chamfered region of the discharge channel upstream the exit plane, which is delimited by a solid horizontal blue line. Section 2 is inside the channel at a location where the magnetic field lines are almost along the axial direction adjacent to the walls. Section 3 is located at a position within the channel where the magnetic flux is almost zero in the central part of the domain along the radius.

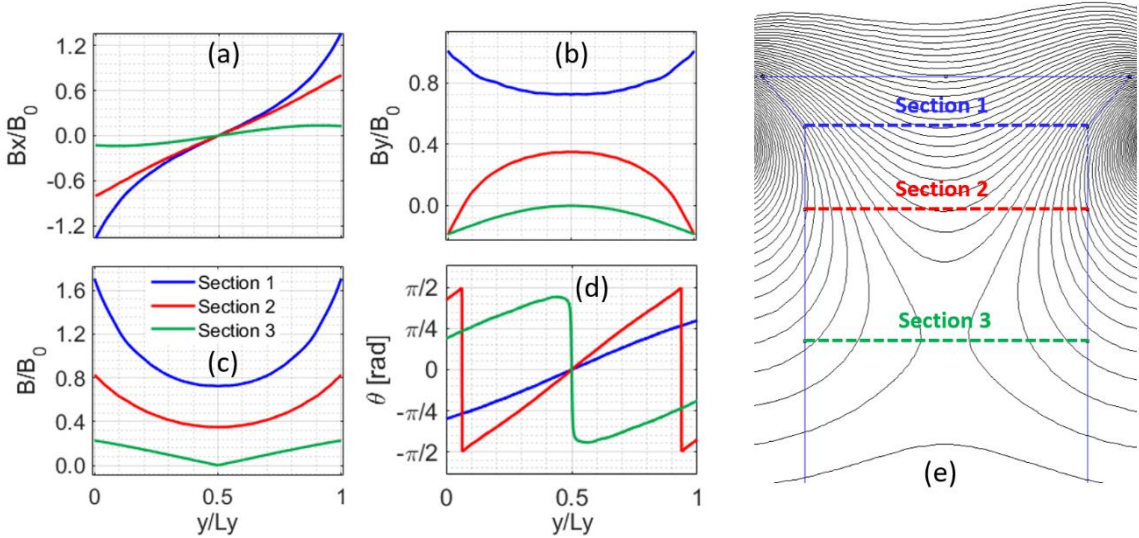

Figure 16: Radial profiles of the magnetic field properties at the three studied Sections; (a) normalized axial component, (b) normalized radial component, (c) normalized magnetic field intensity, and (d) the angle between the tangent to the magnetic field lines and the $x$-coordinate. The normalization factor, $B_0$, is the peak magnetic field intensity for Case II of the axial-azimuthal simulations. In plot (e), the adopted field topology and the locations of the simulated cross-sections are shown.

The operating condition of the HT20k, for which the radial-azimuthal simulations are performed, corresponds to that of Case II of the axial-azimuthal simulations with the anode mass flow rate of 30 mg/s, the discharge voltage of 300 V, and the peak radial magnetic field intensity along the axial direction of $B_0$. As a result, the bulk plasma properties at each simulated cross-section are taken from the time-averaged axial distribution of the properties obtained from the axial-azimuthal simulation of Case II (Figure 9). Additionally, referring to the time evolution plot of the axial electric field for Case II in Figure 8(a) and its time-averaged axial profile in Figure 9(b), it can be noticed that Section 1 falls within the acceleration zone, Section 2 is inside the ionization zone, and Section 3 corresponds to the near-anode zone.

The axial electric field is assumed to be radially constant and is directed along the positive $x$-direction, hence, pointing outward from the simulation plane. For the radial-azimuthal simulations presented in this work, we have neglected the Secondary Electron Emission from the channel walls so as to isolate the effect of the radial magnetic field gradient.

Moreover, as a reference to compare against the results of the simulations with non-uniform radial $B$-field distribution, we have additionally performed, for each cross-section, a quasi-2D simulation with a radially uniform magnetic field distribution whose intensity is equal to the value of the field at the centerline of the channel, i.e., the mid radial location. This set of simulations with uniform radial $B$-field profile also enables us to distinguish the influence on the plasma behavior solely due to the self-consistent resolution of the near-wall plasma sheath and its interactions with the bulk plasma processes.

Table 4 presents a summary of the computational and physical parameters adopted for the quasi-2D simulation of each radial-azimuthal cross-section.

| Parameter | Unit | Value | | |
|---|---|---|---|---|
| | | Section 1 | Section 2 | Section 3 |
| **Computational parameters** | | | | |
| Cell size ($\Delta y = \Delta z$) | $\mu m$ | 20 | | |
| Number of regions along radial direction ($M$) | - | 50 | | |
| Number of regions along azimuthal direction ($N$) | - | 25 | | |
| Time step ($ts$) | s | $2 \times 10^{-12}$ | | |
| Average number of macroparticles per cell for the radial grid ($N_{ppc,y}$) | - | 100 | | |
| Average number of macroparticles per cell for the azimuthal grid ($N_{ppc,z}$) | - | 100 | | |
| Total simulated time | $\mu s$ | 7.5 | | |
| **Physical parameters** | | | | |
| Axial position of the simulated Section ($x/L_x$) | - | 0.88 | 0.68 | 0.36 |
| Initial plasma density ($n_{i,0}$) | $m^{-3}$ | $6.5 \times 10^{17}$ | $1.3 \times 10^{18}$ | $2.35 \times 10^{18}$ |
| Background neutral density ($n_n$) | $m^{-3}$ | $3.5 \times 10^{17}$ | $7 \times 10^{18}$ | $1.3 \times 10^{19}$ |
| Axial electric field ($E_x$) | $Vm^{-1}$ | $1.8 \times 10^4$ | $1 \times 10^3$ | $1 \times 10^2$ |
| Normalized magnetic field at the channel centerline | - | 0.73 | 0.34 | $\approx 0$ |
| Initial electron temperature ($T_{e,0}$) | $eV$ | 20 | 8 | 5 |
| Initial ion temperature ($T_{i,0}$) | $eV$ | 0.5 | | |
| Neutral temperature ($T_n$) | $K$ | 650 | | |

Table 4: Summary of the computational and physical parameters used for the quasi-2D radial-azimuthal simulations

The simulations are initialized by loading the electrons and ions uniformly throughout the simulation domain at an initial number density of $5 \times 10^{17}$ m$^{-3}$ for Section 1, $1.3 \times 10^{18}$ m$^{-3}$ for Section 2, and $2.35 \times 10^{18}$ m$^{-3}$ for Section 3. These initial plasma densities are a factor of 1.2 higher than the plasma density obtained from the axial-azimuthal simulation of Case II at the location of each simulated Section. This is because, as it will be explained shortly, the plasma density in our radial-azimuthal simulations decreases over time. Thus, we increased the initial plasma density so that the average density of the plasma over the duration of the simulation for each cross-section remains close to the corresponding value from the axial-azimuthal simulation. The plasma species are sampled from a Maxwellian at the initial load temperatures reported in Table 4 for each Section.

The simulation domain features a virtual axial length of 1 cm. Along the axial direction, the Poisson's equation is not solved but the particles crossing the boundary in the $x$-direction at $L_x = \pm 1$ cm from the simulation plane are resampled from their initial Maxwellian distribution and are re-introduced onto the simulation plane, maintaining their azimuthal and radial positions. The consideration of such virtual axial extent allows us to limit the growth of the particles' energy in the simulations [45] and is a common practice in the 1D and 2D PIC simulations that do not resolve the axial flux of the particles self-consistently [45][46].

Concerning the particles' boundary condition along the radial direction, the electrons and ions reaching the walls are removed from the simulations. The domain along the azimuth is periodic and, thus, the same periodic boundary condition as that explained in Section 2.1 for the axial-azimuthal simulations are implemented.

The plasma potential in the quasi-2D simulations is solved using the RDPS [24]. Regarding the boundary conditions applied in the potential solver, the radial walls are considered to be electrically floating [47] and the nodes along the two azimuthal ends of the domain are assumed as periodic.

The neutrals are present in the radial-azimuthal simulations as a uniform background with the constant density values provided in Table 4 for each simulated Section. The MCC algorithm [33] is used to resolve the collisions between the electrons and the xenon atoms. The types of collision events considered, and the collision cross-sections used are the same as those mentioned in Section 2.1. An electron undergoing any collision event is isotopically scattered. In the event of an inelastic collision, the particle will experience an energy loss. For the ionization collision, only the associated energy loss is considered for the incident electron and no new plasma particles are created.

As the final point concerning the simulations' setup, we have followed the approach suggested in Ref. [48] to compensate for the radial flux of particles lost to the walls. This is necessary because the simulations do not self-consistently resolve the axial inflows and outflows of the particles. The radial fluxes are compensated by injecting electron-ion pairs randomly in the domain at each timestep. The number of pairs to inject is estimated as the minimum between the number of electrons and the number of ions reaching the wall at every timestep [48]. It is noteworthy that the adopted approach only partially compensates for the particles lost to the wall, which implies that the plasma density gradually decreases over time. In this regard, we did not choose to impose a plasma source to maintain the plasma density constant throughout the simulations because the impacts of such an artificial source cannot be readily assessed in comparison with allowing the ionization collisions to produce plasma self-consistently. Accordingly, we run the radial-azimuthal simulations for 7.5 $\mu$s only so that we ensure that, across all simulations and for each simulated cross-section, the radially averaged plasma density remains on average almost the same as the time-averaged plasma density along the channel centerline from the axial-azimuthal simulation at that cross-section.

### 3.2. Results and discussion

Figure 17 shows the time evolution of the spatially averaged ion (plasma) density from the radial-azimuthal simulations. The dotted curves correspond to the simulations with uniform radial distributions of the magnetic field intensity whereas the solid curves are from the simulations with non-uniform radial profiles of $B$-field.

It is evident that, overall, the ion density for all simulated Sections decreases over time. However, the rate of decrease in $n_i$ is different across various Sections and also varies depending on the radial distribution of the $B$-field in each Section. In this regard, at Section 3, the simulated case with uniform radial $B$ profile (dotted green curve) almost corresponds to an electrostatic problem due to very low magnetic field intensities and, therefore, the $n_i$ decreases quite rapidly. Nonetheless, in case of non-uniform $B$-field distribution for Section 3, as seen in Figure 16(c), the rate of decrease in the ion density is reduced as the existing $\nabla B$ force introduces some degree of plasma confinement. In this respect, as the radial gradient of the magnetic field intensity and, thus, the grad-$B$ force increases from Section 3 to Section 1, we notice that for quasi-2D simulations of Section 1 and 2 with non-uniform $B$-field profile, the rate of density decrease is significantly reduced, and the density almost approaches an asymptotic value for these Sections from about 2 $\mu s$ into the simulations.

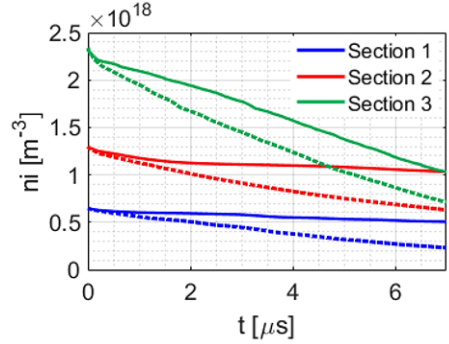

Figure 17: Time evolution of the ion number density from the quasi-2D simulations of the three studied Sections with non-uniform (solid curves) and uniform (dotted curves) radial magnetic field distributions. For the simulations with radially uniform $B$ profile, the normalized magnetic field values ($B/B_0$) are 0.73 for Section 1, 0.34 for Section 2, and approximately 0 for Section 3.

The confinement of the plasma at Sections 1 and 2 can be also noticed from Figure 18(a) by comparing the corresponding time-averaged density profiles for each Section from the quasi-2D simulations with non-uniform and uniform radial distributions of the $B$-field. We observe that, due to the radial magnetic field gradient, the ion density in the center of the domain is strongly increased at both Sections 1 and 2, with the confinement being slightly stronger at Section 2. At Section 3, we do not observe a concentration of the plasma in the center of the domain similar to that for Sections 1 and 2. Nevertheless, in the case with non-uniform radial $B$-field distribution, the plasma density at Section 3 remains almost constant over a larger radial extent, which implies that the length scale of the plasma sheath is lower in this case compared to the case with uniform radial $B$-field profile.

Looking at Figure 18(b), the time-averaged $T_e$ profile of Section 1 from the simulation with uniform $B$ profile (dotted blue curve) is overall higher than the corresponding profile from the simulation with non-uniform $B$ profile. Nevertheless, the $T_e$ around the mid radial location at Section 1 is about 60 eV in either case. Concerning Sections 2 and 3, the $T_e$ values at Section 3, which represents a cross-section in the near-anode zone, are observed to be interestingly larger than those for Section 2 with either uniform or non-uniform radial $B$ profiles. The reason for this observation can be traced to the fact that, at Section 3, the magnetic field intensity is zero at the center of domain. The overall magnitude of the magnetic field is also quite low at this Section. As a result, the consequent de-magnetization of the electrons may have led to their thermalization, and thus higher temperatures compared to Section 2.

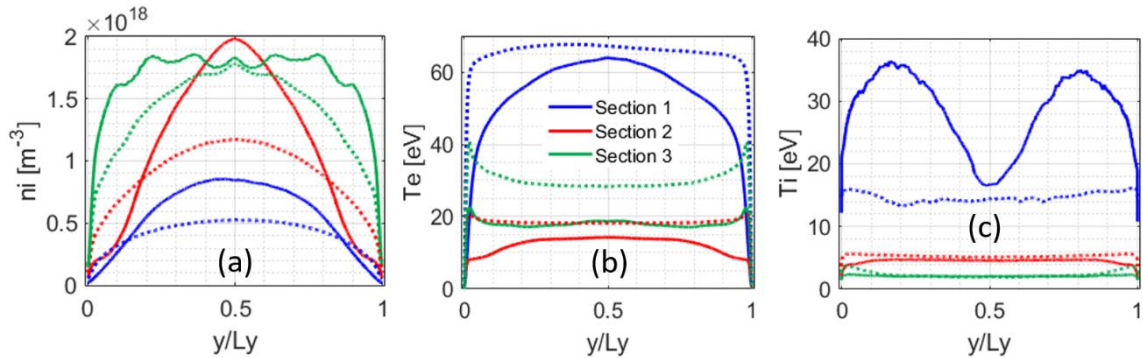

Figure 18: Time-averaged (over 7 $\mu s$) radial profiles of the plasma properties for the three simulated Sections; (a) ion number density, (b) electron temperature, (c) ion temperature. The dotted curves correspond to the simulations with uniform radial magnetic field distribution, and the solid curves are from the simulations with non-uniform radial $B$ profile.

Referring to the $T_e$ profiles in Figure 18(b), we acknowledge that the predicted electron temperatures from our simulations might be, to a certain extent, higher than that occurring in the real-world and that predicted by other 2D hybrid simulations of magnetically shielded thrusters [21][23] since we have neglected the effect of the Secondary Electron Emission from the channel walls, which can lower the average temperature of the electron population.

Figure 18(c) shows that the time-averaged ion temperature profile at Section 1 features a much stronger radial variation and higher temperature values from the simulation with non-uniform radial $B$-field distribution compared to the $T_i$ profile from the simulation case with uniform $B$ profile. As we will see later in this section, the elevated $T_i$ values at Section 1 with the non-uniform radial $B$ profile are due to the interactions of the ion particles with a long-wavelength azimuthal wave mode that has developed at this Section. The $T_i$ values at Section 2 are seen to be consistently higher with respect to those at Section 3 regardless of the radial $B$ profile.

An important point to highlight here is that the electron and ion temperatures at all three Sections are increased compared to their initial values that were derived from the results of the axial-azimuthal simulations. Moreover, because of the radial $B$-field gradient, the profiles of the plasma density and the species' temperature exhibit notable radial gradients, particularly at Sections 1 and 2 that correspond to cross-sections, respectively, within the acceleration and ionization zone. Accordingly, as anticipated in Section 2.2, the axial dynamics of the discharge and the plasma processes such as the ionization are expected to be largely influenced by these radial gradients. This once again emphasizes the highly 3D nature of the plasma configuration in magnetically shielded Hall thrusters.

The main purpose of the magnetic shielding topology is to mitigate the power deposition mainly from the ions on the channel walls, particularly around the exit plane, so as to lower the erosion rate of the ceramic channel and, hence, to extend the thruster's lifetime [49]. Accordingly, we present, in Figure 19, the time evolution plots of the electron and ion power deposition on the channel walls for various simulated Sections. In Figure 19, the plots on the first row are from the simulations with non-uniform radial $B$-field profile, whereas the plots on the second row correspond to the simulations with uniform radial $B$ distribution. Comparing the time-averaged $P_i$ values from the simulations with non-uniform and uniform radial $B$ profiles (Figure 19(b) and (d)), we notice that the radial gradient of the magnetic field associated with the shielding topology has indeed reduced the power deposition at Sections 1 and 2 by a factor of about 4 and 8, respectively. Moreover, the time-averaged $P_i$ at Section 3 is lowered by a factor of 1.5 due to the radial field gradient. Similarly, referring to Figure 19(a) and (c), the mean $P_e$ values are also observed to be lowered as a result of the magnetic shielding topology.

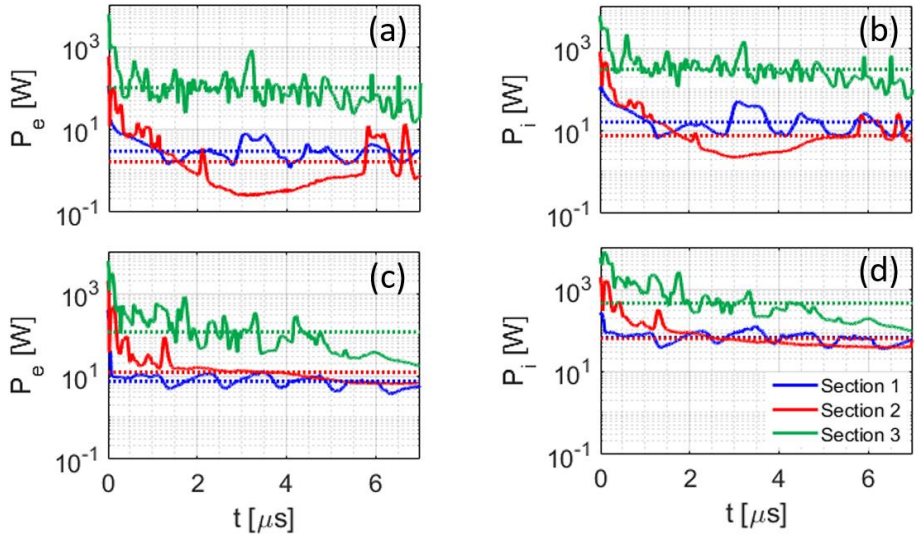

Figure 19: Time evolution of the average total power deposition on the radial walls from the quasi-2D simulations of the three Sections; (a) and (c) electron power deposition, (b) and (d) ion power deposition. The plots on the top row are from simulations with non-uniform radial magnetic field distribution and the plots on bottom row correspond to simulations with uniform radial $B$ field.

The oscillations in the time evolution plots of $P_e$ and $P_i$ are because the particles' energy varies over time due to their interactions with the azimuthal instabilities. In this regard, for Section 1, the frequency of the oscillations in $P_e$ and $P_i$ is very similar regardless of the radial field distribution whereas the oscillations' amplitude is slightly larger for the case with non-uniform radial $B$ profile. Concerning Section 2, we notice that the temporal variations in $P_e$ and $P_i$ between the simulation cases with non-uniform and uniform radial $B$ profiles are notably different.

Also noteworthy is the fact that the temporal decay in the electron and ion power deposition, particularly observed for the simulations with uniform radial $B$ profile (Figure 19(c) and (d)), is associated with the gradual decrease in the plasma density itself as seen in Figure 17.

The time-averaged radial profiles of the electrons' axial velocity ($V_x$) and axial mobility ($\mu_x$) for the three simulated Sections are shown in Figure 20. The axial velocity profiles (Figure 20(a)) are inversed by multiplying the velocity values by -1. The axial mobility distributions in Figure 20(b) are obtained by dividing the time-averaged velocity profile of each Section by the axial electric field value at that Section. Starting with Section 1, the axial velocity of the electrons exhibits a strong variation along the radius from the simulation case with non-uniform radial $B$ distribution (solid blue curve). Moreover, the overall electron axial velocity at Section 1 is the highest among the three Sections regardless of the radial $B$ profile. In any case, as the axial electric field is also the largest at Section 1, the mean axial mobility at this Section with the value of 40 $m^2V^{-1}s^{-1}$ is the lowest among the three Sections.

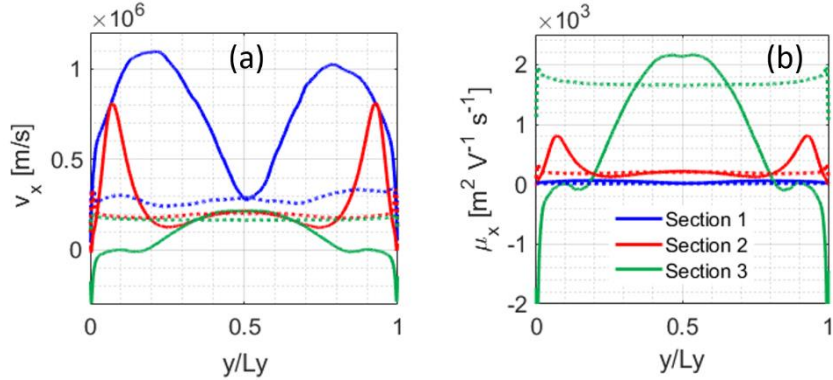

Figure 20: Time-averaged (over 7 $\mu s$) radial profiles of (a) electron axial velocity, and (b) electrons' axial mobility, for the three simulated Sections. The solid curves represent simulations with non-uniform magnetic field distribution, and the dotted curves correspond to simulations with uniform radial $B$ field.

Concerning Section 2, we notice that the electrons' axial velocity and axial mobility both increase toward the walls. This is because, at Section 2, the magnetic field lines near the walls are almost along the axial direction (Figure 16(e)), which implies that the electrons gyrating around these lines move axially in the direction opposite to the axial electric field, hence, showing an increased the axial velocity and mobility.

At Section 3, it is observed that the electrons funnel out of the simulation plane from the central part of the domain where the magnetic flux is almost zero (Figure 16). In this regard, in the case of uniform radial $B$ profile, the axial velocity and mobility of the electrons are almost radially constant, whereas in the case of non-uniform radial $B$ distribution, the axial velocity and mobility profiles show a notable radial variation. Particularly, the electrons are seen to move in the direction of the axial electric field as we approach the radial ends of the domain at either side.

We now delve into the characteristics and the evolution of the azimuthal instabilities from the radial-azimuthal simulations. To this end, we have shown, respectively, in plots (a) and (b) of Figure 21 the 1D temporal and the 1D spatial FFTs of the azimuthal electric field signal from the quasi-2D simulations of each Section. We observe that, at Section 1, the dominant frequency and wavenumber content of the azimuthal electric field fluctuations shift toward lower-frequency, longer wavelength modes in the presence of the radial magnetic field gradient (solid blue curves). Indeed, in the case of uniform radial $B$ profile (dotted blue curves), the dominant frequency content is broader and ranges from 4-12 MHz, whereas the dominant frequencies are concentrated around the values of 1-1.5 MHz in the case of non-uniform radial $B$ profile. Similarly, the dominant azimuthal wavenumber range has changed from 1-2 mm in the case of uniform radial $B$ profile to 9-18 mm in the presence of the radial gradient of the $B$-field.

It is interesting to note that the frequency range of the dominant azimuthal modes in the absence of the radial $B$-field gradient is similar to the range observed from the axial-azimuthal simulation of Case II in Figure 11.

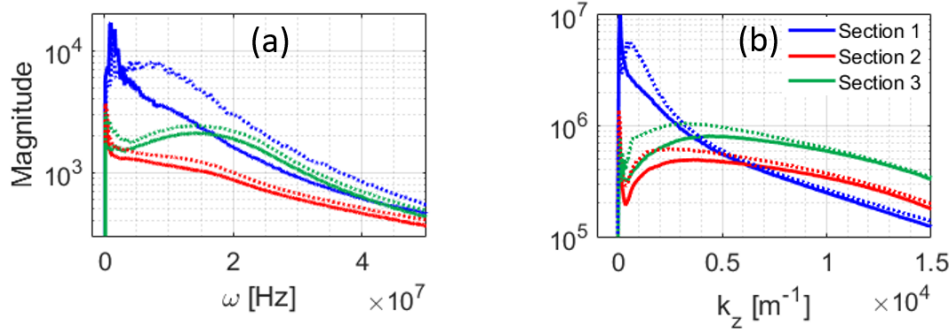

Figure 21: (a) 1D temporal, and (b) 1D spatial FFT plots of the azimuthal electric field signal, averaged over all radial positions, for the three simulated Sections. The spatial FFTs are also averaged over 7 $\mu$s of the simulations' time. The temporal FFTs correspond to the electric field fluctuations at mid azimuthal location. The solid curves represent simulations with non-uniform radial $B$ profile, and the dotted curves correspond to simulations with uniform radial $B$ profile.

Regarding Section 2, it is seen that that radial magnetic field gradient does not have a notable impact on the frequency and azimuthal wavenumber of the dominant modes. Indeed, either in the absence or presence of the radial $B$-field gradient, the frequency spectrum is dominated by modes with very low frequencies. The wavenumber spectrum of Section 2 shows a distinct peak at very low azimuthal wavenumbers and also a broad peak toward higher wavenumbers.

The frequency and wavenumber spectra of the azimuthal wave modes at Section 3 are rather similar to those of Section 2. Nevertheless, the overall FFT magnitude of the modes are higher for Section 3, and, in addition, the frequency spectrum also shows a broad secondary peak in the frequency range of 10-25 MHz.

As Section 1 lies within the acceleration zone, we take a closer look at the dynamics of the azimuthal modes at this section in the presence and absence of the radial magnetic field gradient to cast light on the impact of the magnetic shielding topology on the evolution of the azimuthal waves around the channel exit plane. In this respect, Figure 22 shows the time evolution of the 1D spatial FFTs of the axial electron current ($J_{ex}$) signal, averaged over all radial positions. The approach to obtain the plots in Figure 22 are similar to that explained in Ref. [50].

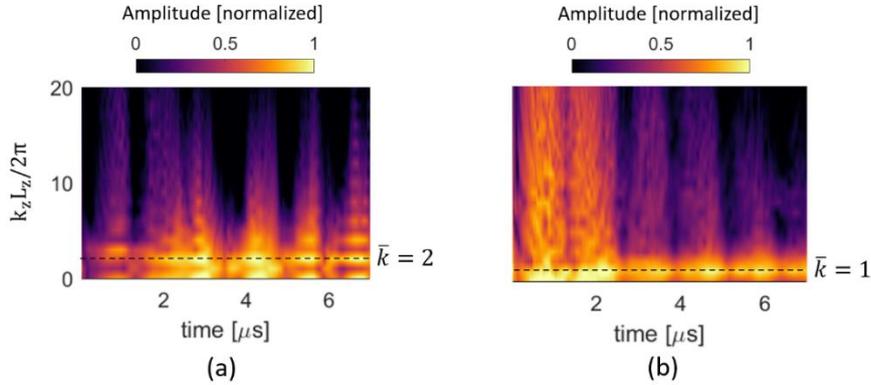

Figure 22: The time evolution of 1D spatial FFTs of the $J_{ex}$ signal from the simulations of Section 1 with (a) non-uniform and, (b) uniform radial $B$ profiles. $\bar{k}$ indicates the normalized primary azimuthal wavenumber in each case.

It is interesting to note that, in the case of non-uniform radial $B$ profile, the evolution of the modes in Figure 22(a) shows a behavior resembling the inverse energy cascade [50][51] with the spectrum of the waves periodically narrowing toward longer wavelengths. In addition, particularly from 5 $\mu$s until the end of the simulation, we see the formation of the harmonics of the azimuthal waves with the normalized primary azimuthal wavenumber ($\bar{k}$) of 2.

In case of uniform radial $B$ profile, (Figure 22(b)), we can again notice the inverse energy cascade toward the longer-wavelength modes. However, no distinct harmonics of the azimuthal waves are observed, and the spectrum is quite broadband. It is also important to note that the apparent decrease in the magnitude of the modes over time in Figure 22(b) is because of the gradual plasma density decrease as seen in Figure 17.

The dynamics of the azimuthal waves at Section 1 in the presence and absence of the radial $B$-field gradient is shown in Figure 23 in terms of the cyclic evolution of the 2D snapshots of the $J_{ex}$ from the quasi-2D simulations.

Also, Figure 24 shows the 2D snapshots of various plasma properties from the simulations with non-uniform and uniform radial $B$-fields at a time corresponding to the beginning of the evolution cycles in Figure 23(a) and (b), i.e., the upper left-most subplots.

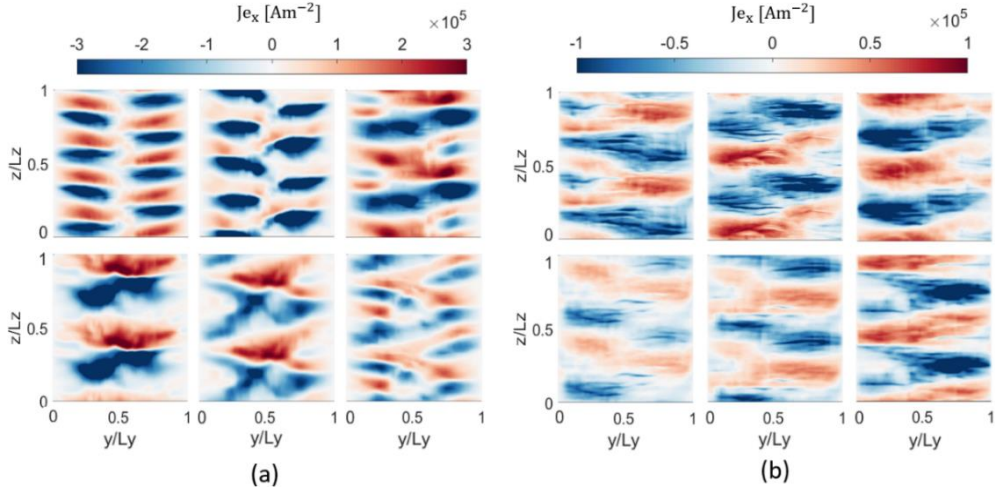

Figure 23: 2D snapshots of the axial electron current density ($J_{ey}$) at six different moments through the discharge evolution at Section 1 with (a) non-uniform, and (b) uniform radial $B$-field. The time advances from left to right at each row.

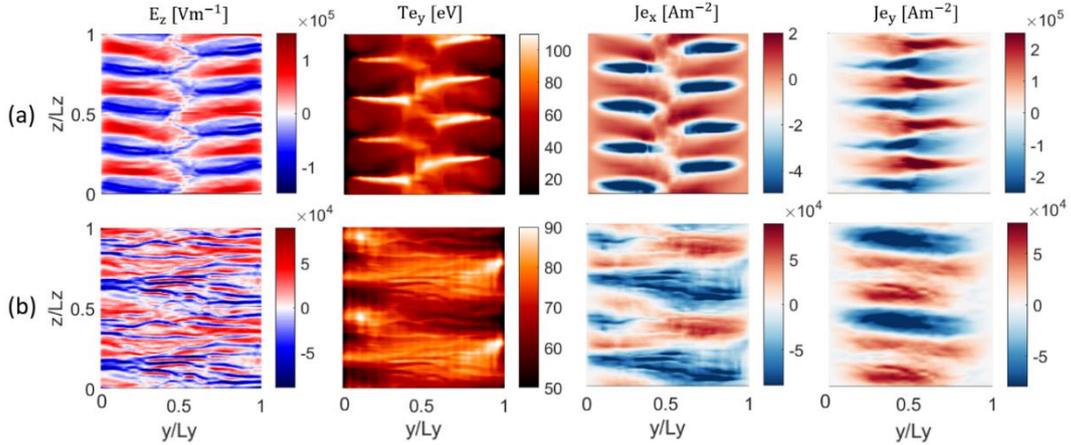

Figure 24: 2D snapshots of the plasma properties at Section 1 with (a) non-uniform, and (b) uniform radial $B$-field. The columns, from left to right, represent the azimuthal electric field ($E_z$), radial electron temperature ($T_{ey}$), and the axial and radial electron current densities ($J_{ex}$ and $J_{ey}$).

From Figure 23(a), we observe that, in the presence of radial $B$-field gradient, the azimuthal waves initially feature both azimuthal and radial wavenumbers. The azimuthal and radial wavenumbers of the modes at the beginning of the discharge evolution cycle is also evident from the 2D snapshots of the azimuthal electric field and the radial electron temperature in Figure 24(a). However, the 2D distribution of the radial electron current density ($J_{ey}$) in the right-hand side of Figure 24(a) mostly features the azimuthal wavenumber. Indeed, along the radius, $J_{ey}$ is more focused around the center of the domain and becomes rather dispersed toward the walls. In any case, as the discharge evolves in Figure 23(a), the waves with azimuthal and radial wavenumber components are seen to merge and a long-wavelength mode with only an azimuthal wavenumber is formed. This mode subsequently breaks away and the initial azimuthal waves begin to reappear and become dominant.

In contrast, in the case of uniform radial $B$-field, it is observed from plot (b) in Figure 23 and Figure 24 that a spectrum of longer- and shorter-wavelength azimuthal modes are present in the 2D snapshots of all plasma properties and the "turbulent" nature of the plasma persists over time. In this regard, from Figure 23(b), certain wave modes seem to periodically mitigate and grow as the discharge evolves, but no mode appears to become distinctly dominant through the discharge evolution.

The normalized velocity distribution functions of the electrons and the ions along the axial and azimuthal velocity components are presented in Figure 25 for each of the simulated Sections. For each Section, the VDFs are shown from the simulations with non-uniform and uniform radial $B$-field profiles.

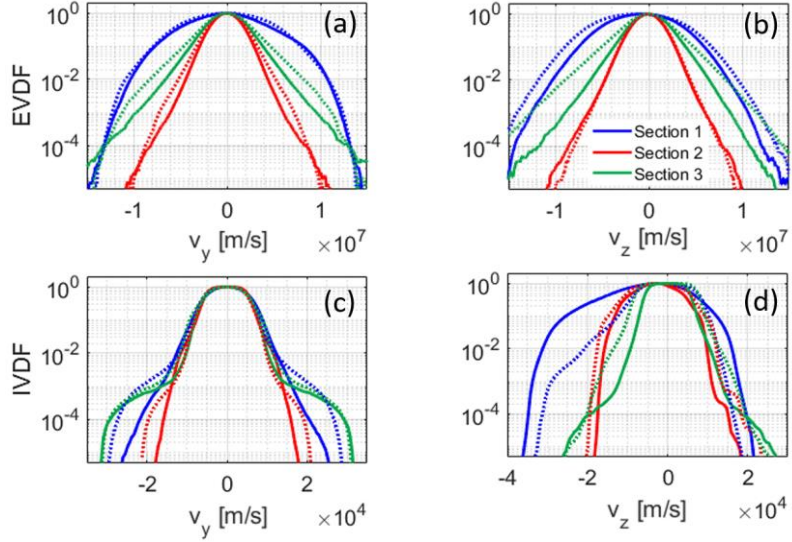

Figure 25: Normalized VDFs at t = 3 $\mu s$ for the three simulated Sections. Top row: the electrons' VDF along (a) the radial and (b) the azimuthal velocity component; Bottom row: the ions' VDF along (a) the radial and (b) the azimuthal velocity component. The solid curves represent simulations with non-uniform radial $B$ profile, and the dotted curves correspond to simulations with uniform radial $B$ profile.

Referring to plots (a) and (b) in Figure 25, the radial VDF of the electrons shows minor to no effect from the radial magnetic field gradient across all three Sections. However, at Sections 1 and 3, the azimuthal VDF shows some degree of broadening in the case of uniform radial $B$ profile, which is more pronounced at Section 3. Additionally, as it has been observed in previous investigations of the effect of the radial magnetic field gradient on the plasma processes [29][52], the electrons' VDF at all three Sections is seen to be rather isotropic. This isotropization is most evident for Section 2 where the EVDF is interestingly also isotropic in the case of uniform radial $B$ profile.

The ions' radial VDF in plot (c) of Figure 25 shows a broadening of the distribution function at Sections 1 and 2 in the case of uniform radial $B$ profile from the radial velocities slightly higher than $\pm$ 10 km/s. At Section 3, the radial IVDF has notable side lobes in either case of uniform or non-uniform radial $B$-field profile.

Plot (d) in Figure 25 illustrates a distortion of the azimuthal IVDF at all three Sections. In particular, the azimuthal waves observed to be present at Section 1 are noticed to have caused a significant heating of the ion population, which is manifested as a notable broadening of and asymmetry in the IVDF. The ion heating is stronger in the case of non-uniform radial $B$ profile where a long-wavelength, low-frequency azimuthal wave mode was seen to develop and become dominant.

The ions heating along the radial and azimuthal directions at Section 3 is believed to be the result of the excitation of waves whose characteristics are affected by the almost de-magnetization of the electrons at this Section.

**Section 4: Conclusions**

In this article, we presented an extensive set of results from the quasi-2D PIC simulations of the HT20k Hall thruster in the axial-azimuthal and radial-azimuthal configurations. The results and analyses provided in this work underline the flexibility and the applicability of the reduced-order PIC scheme to serve as a reliable, predictive numerical tool to not only study the underlying multi-dimensional plasma processes and interactions but also to potentially aid the development of advanced Hall thrusters. The numerous simulations presented are made possible by the cost-efficiency of the reduced-order PIC scheme that enables performing self-consistent and kinetic studies in real-size Hall thrusters.

The axial-azimuthal simulations were carried out for a rather simplified representation of the real-world thruster in three operating conditions using the single-region approximation of the 2D problem. In the simplified axial-azimuthal representation, the simulations cannot resolve the radial gradients in the magnetic field and the plasma properties that are inherent to the magnetic shielding topology. The main purpose of these simulations was, hence, to evaluate the utility of the single-region quasi-2D simulation for reliable prediction of the plasma dynamics and global performance parameters in Hall thrusters. In this respect, we observed that the axially averaged instability-induced electron mobility resolved by the single-region simulations enabled a self-consistent prediction of the plasma discharge behavior in HT20k over a range of operating conditions for an unprecedented duration of 650 $\mu s$ without the need to use any ad-hoc electron transport model.

Moreover, considering the simplified problem description adopted and the approximation order used for the axial-azimuthal simulations, the predicted performance parameters were found to have a reasonable maximum error of 20-30 % across the three simulated operating conditions. It was underlined that the main source of discrepancy between the simulated and the measured performance is the lack of resolving the important effect of the radial magnetic field gradient in the axial-azimuthal simulations. This justification was confirmed by the results of the HT20k radial-azimuthal simulations which showed that the radial gradients have a notable influence on the underlying plasma phenomena and the macroscopic properties in magnetically shielded Hall thrusters.

Nonetheless, the axial-azimuthal simulations also provided some insights into the average characteristics of the azimuthal instabilities in HT20k. The waves' characteristics were noticed to be consistent with the theoretical dispersion relation of the ion acoustic waves. In addition, the instabilities' dispersion map was seen to show a continuous spectrum with no particularly dominant mode. The numerical dispersion of the waves was observed to change between two rather distinct states manifested as a bifurcation in the dispersion plots of the azimuthal instabilities.

The radial-azimuthal quasi-2D simulations were performed using a high-fidelity multi-region approximation of the problem at three cross-sections of the thruster's discharge channel. The simulations with uniform and non-uniform radial profiles of the magnetic field showed that, in the presence of the radial magnetic field gradient, the plasma becomes concentrated in the center of the domain away from the radial walls. Moreover, we observed that, in the case of non-uniform radial $B$ profile associated with the magnetic shielding topology, the power deposition from the plasma species on the channel walls is reduced by several factors, particularly at the Sections inside the channel and around the exit plane.

The evolution and the characteristics of the azimuthal instabilities were also noticed to be majorly affected by the radial $B$-field gradient. In particular, at a Section within the acceleration zone, a long-wavelength, relatively low-frequency wave mode was seen to develop that induces a significant electrons' axial mobility and a notable heating of the ion population.

It is important to note that the simulations' results from this work have not been directly compared against detailed experimental measurements, such as the distribution of the intensive plasma properties, because such data were unavailable for the HT20k. Accordingly, the insights derived from the axial-azimuthal and radial-azimuthal simulations, which were summarized in the preceding paragraphs, shall be viewed in light of this absence of an experimental validation.

**Acknowledgments**:

The present research is carried out within the framework of the project "Advanced Space Propulsion for Innovative Realization of space Exploration (ASPIRE)". ASPIRE has received funding from the European Union's Horizon 2020 Research and Innovation Programme under the Grant Agreement No. 101004366. The views expressed herein can in no way be taken as to reflect an official opinion of the Commission of the European Union.

MR, FF, and AK gratefully acknowledge the computational resources and support provided by the Imperial College Research Computing Service (http://doi.org/10.14469/hpc/2232).

**Data Availability Statement**:

The simulation data that support the findings of this study are available from the corresponding author upon reasonable request. The HT20k thruster's information and test data, used with permission for the present work, are proprietary of SITAEL SpA and are not available to the public.